\documentclass[apj]{emulateapj}
\usepackage{graphicx}

\def\simgr{\,\hbox{\hbox{$ > $}\kern -0.8em \lower 1.0ex\hbox{$\sim$}}\,}
\def\simle{\,\hbox{\hbox{$ < $}\kern -0.8em \lower 1.0ex\hbox{$\sim$}}\,}

\newcommand\rxone{1RXS J154439.4$-$112820}
\newcommand\xss{XSS J12270$-$4859}
\newcommand\fglone{3FGL J1544.6$-$1125}
\newcommand{\rxs}{1RXS J083842.1$-$282723}
\newcommand{\msp}{XMMU J083850.38$-$282756.8}
\newcommand{\qso}{XMMU J083842.85$-$282831.8}
\newcommand{\fgl}{3FGL J0838.8$-$2829}
\newcommand{\fermi}{{\it Fermi}}
\def\rosat{{\it ROSAT\/}}
\def\chandra{{\it Chandra}}
\def\xmm{{\it XMM--Newton}}

\def\spin{$94.8\pm0.4$ minutes}
\def\orbit{$98.3\pm0.5$ minutes}
\def\beat{$14.7\pm1.2$~hr}
\def\optper{$98.413\pm0.004$~minutes}

\slugcomment{}
\received{2016 November 8}
\accepted{2017 January 13}
\shortauthors{Halpern et al.}
\shorttitle{Observations of 3FGL J0838.8$-$2829}

\begin{document}
\title{X-ray and Optical Study of the Gamma-ray Source 3FGL J0838.8$-$2829:
Identification of a Candidate Millisecond Pulsar Binary \\
and an Asynchronous Polar}

\author{Jules P. Halpern\altaffilmark{1}, Slavko Bogdanov\altaffilmark{1},
and John R. Thorstensen\altaffilmark{2}} 
\altaffiltext{1}{Columbia Astrophysics Laboratory, Columbia University,
550 West 120th Street, New York, NY 10027-6601; jules@astro.columbia.edu}
\altaffiltext{2}{Department of Physics and Astronomy, 6127 Wilder Laboratory,
Dartmouth College, Hanover, NH 03755-3528}

\begin{abstract}
We observed the field of the \fermi\ source \fgl\ in optical and X-rays,
initially motivated by the cataclysmic variable (CV) \rxs\
that lies within its error circle.  Several X-ray sources first classified
as CVs have turned out to be $\gamma$-ray emitting millisecond pulsars (MSPs).
We find that \rxs\ is in fact an unusual CV, a stream-fed
asynchronous polar in which accretion switches between magnetic poles
(that are $\approx120^{\circ}$ apart) when the accretion rate is
at minimum.  High-amplitude X-ray modulation at periods of \spin\
and \beat\ are seen.  The former appears to be the spin period,
while the latter is inferred to be one-third of the beat period
between the spin and the orbit, implying an orbital period of \orbit.
We also measure an optical emission-line spectroscopic period of \optper,
which is consistent with the orbital period inferred from the X-rays.
In any case, this system
is unlikely to be the $\gamma$-ray source.  Instead, we find a fainter
variable X-ray and optical source, \msp, that is modulated on a time scale
of hours in addition to exhibiting occasional sharp flares.
It resembles the black widow or redback pulsars that have been discovered
as counterparts of \fermi\ sources, with the optical modulation
due to heating of the photosphere of a low-mass companion star by,
in this case, an as-yet undetected MSP.  We propose \msp\ as the MSP
counterpart of \fgl.
\end{abstract}


\section{Introduction}

The Large Area Telescope on the \fermi\ Gamma-ray Observatory
has detected numerous young pulsars, as well as recycled millisecond
pulsars (MSPs) in close binary systems.  Most prominent of the new
discoveries are the black widow (BW) pulsars and so-called
``redback'' systems \citep{rob13},
which comprise a large fraction of the MSPs selected by \fermi. 
The BWs are MSPs with sub-stellar mass, degenerate companions,
while the redbacks generally have $>0.1\,M_{\odot}$ evolved companions.
The latter are usually close to filling their Roche-lobes,
which makes them a direct link to the low-mass X-ray
binary (LMXB) progenitors of MSPs.

Recently, three redbacks have been observed to transition between
radio pulsar and accreting states on timescales of years:
PSR J1023+0038 \citep{arc09}, \xss\ \citep{roy15}, and PSR J1824$-$2452I
in the globular cluster M28 \citep{pap13}.  All of these are hard X-ray
and/or $\gamma$-ray sources.  PSR J1023+0038 was initially misclassified
as a cataclysmic variable (CV) (\citealt{bon02}, but see \citealt{tho05}
for a contrary interpretation),
as was \xss\ \citep{mas06,but08}, because of the similarity of
their optical emission-line spectra and luminosities to those
of CVs.  Now we know that transitional MSPs are distinguishable from CVs
by their X-ray and optical light curves, which show characteristic
dips and flares that are unique to this class \citep{bog15a,dem13}.
We employed this test to reevaluate two \rosat\
All-Sky Survey X-ray sources in \fermi\ error circles that were
spectroscopically classified as CVs by \citet{mas13}.
Using X-ray and optical time-series data for one of these,
\rxone/\fglone, we concluded that it is almost
certainly an MSP binary in the accreting state \citep{bog15b}.
Here, we report on a similar investigation of the second Masetti et al.
CV, \rxs\ in the error circle of \fgl\ \citep{ace15}.

Section 2 describes the observations obtained.
Section 3 presents the results of optical and X-ray observations of
\rxs, which show that it is indeed a CV, probably an
asynchronous polar (AM Herculis star).  Section 4 reports the
discovery of a second X-ray and optical source in the \fermi\ error circle,
\msp, which we identify as a candidate BW pulsar system, and likely
the counterpart of the $\gamma$-ray source.  In Section 5 we show that
a third X-ray source in the \fermi\ error circle is a QSO, probably
unrelated to the $\gamma$-ray source.
Section 6 discusses the properties of the CV and the MSP candidate in
relation to other objects in their respective classes.

\begin{deluxetable*}{lccccl}
\tablewidth{0pt}
\tablecaption{Log of MDM Observatory Time-Series Photometry of \rxs}
\tablehead{
\colhead{Telescope/Detector} & \colhead{Date (UT)} & \colhead{Time (TDB)} &
\colhead{Filter} & \colhead{Exposure (s)} & \colhead{Conditions}
}
\startdata
2.4~m/Templeton & 2014 Mar 22 & 02:47--06:47 & V     & 10 & Clear \\
2.4~m/Templeton & 2014 Mar 23 & 02:40--06:51 & V     & 10 & Photometric \\
1.3~m/Templeton & 2015 Feb 17 & 06:12--09:19 & BG38  & 20 & Partly cloudy \\
1.3~m/Templeton & 2015 Feb 18 & 03:51--09:14 & BG38  & 20 & Photometric \\
1.3~m/Andor     & 2016 Mar 16 & 04:31--06:31 & GG420 & 5  & Photometric \\
1.3~m/Andor     & 2016 Mar 17 & 04:02--06:02 & GG420 & 5  & Photometric 
\enddata
\label{tab:optlog}
\end{deluxetable*}

\begin{deluxetable}{clcc}
\tabletypesize{\small}
\tablewidth{0pt}
\tablecaption{Optical Positions}
\tablehead{
\colhead{Label} & \colhead{Source} & \colhead{R.A. (h m s)} &
\colhead{Decl. ($^{\circ\ \prime\ \prime\prime}$)}
}
\startdata
a  & \rxs\  &  08 38 43.34  &  --28 27 00.9 \\
b  & \msp\  &  08 38 50.45  &  --28 27 57.4 \\
c  & \qso\  &  08 38 42.80  &  --28 28 31.0
\enddata
\tablecomments{Coordinates are equinox J2000.0}
\label{tab:positions}
\end{deluxetable}

\begin{figure*}
\vspace{-0.8in}
\centerline{
\includegraphics[angle=0,width=0.56\linewidth,clip=]{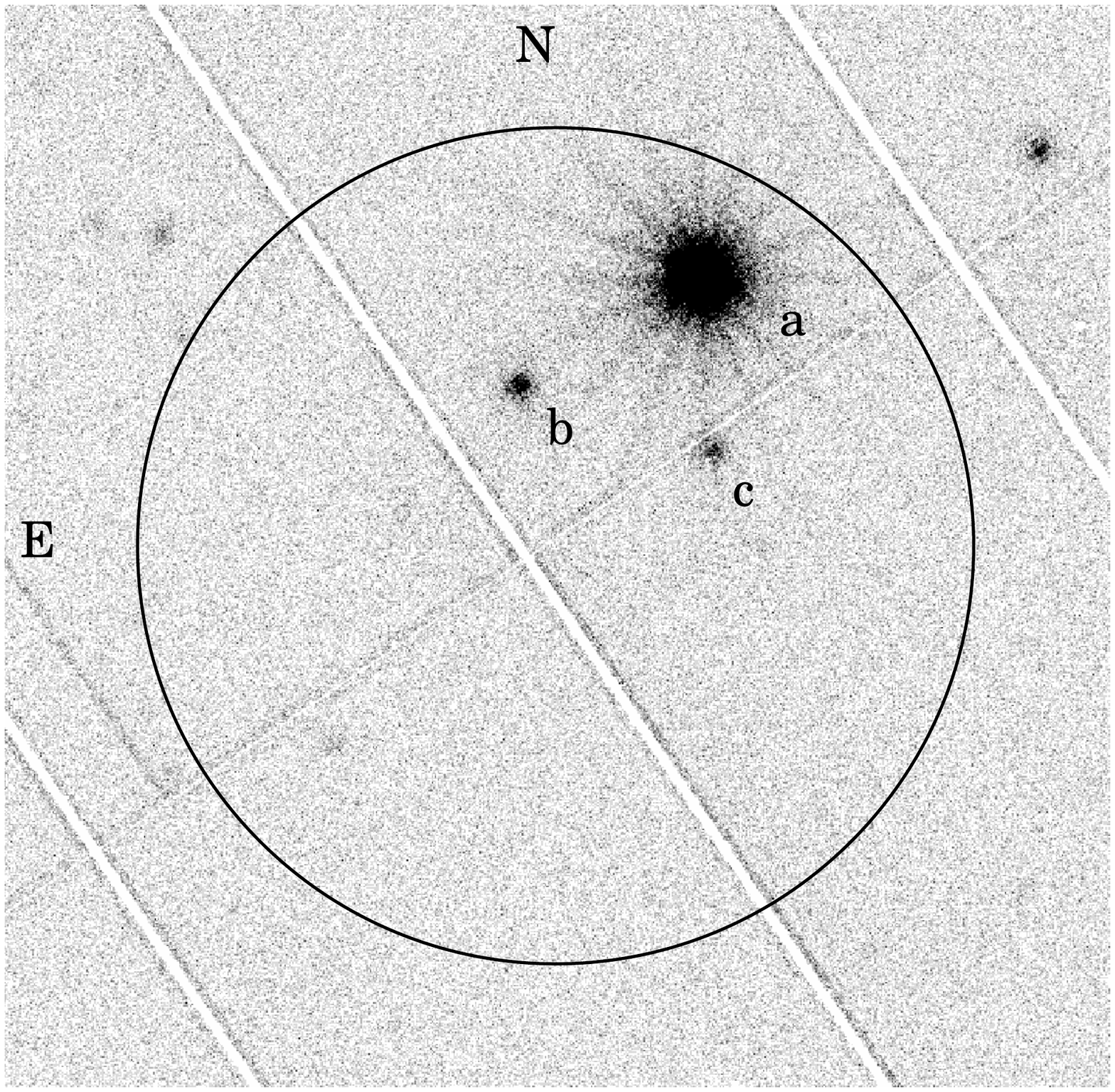}
\hspace{-0.4 truein}
\includegraphics[angle=0,width=0.55\linewidth,clip=]{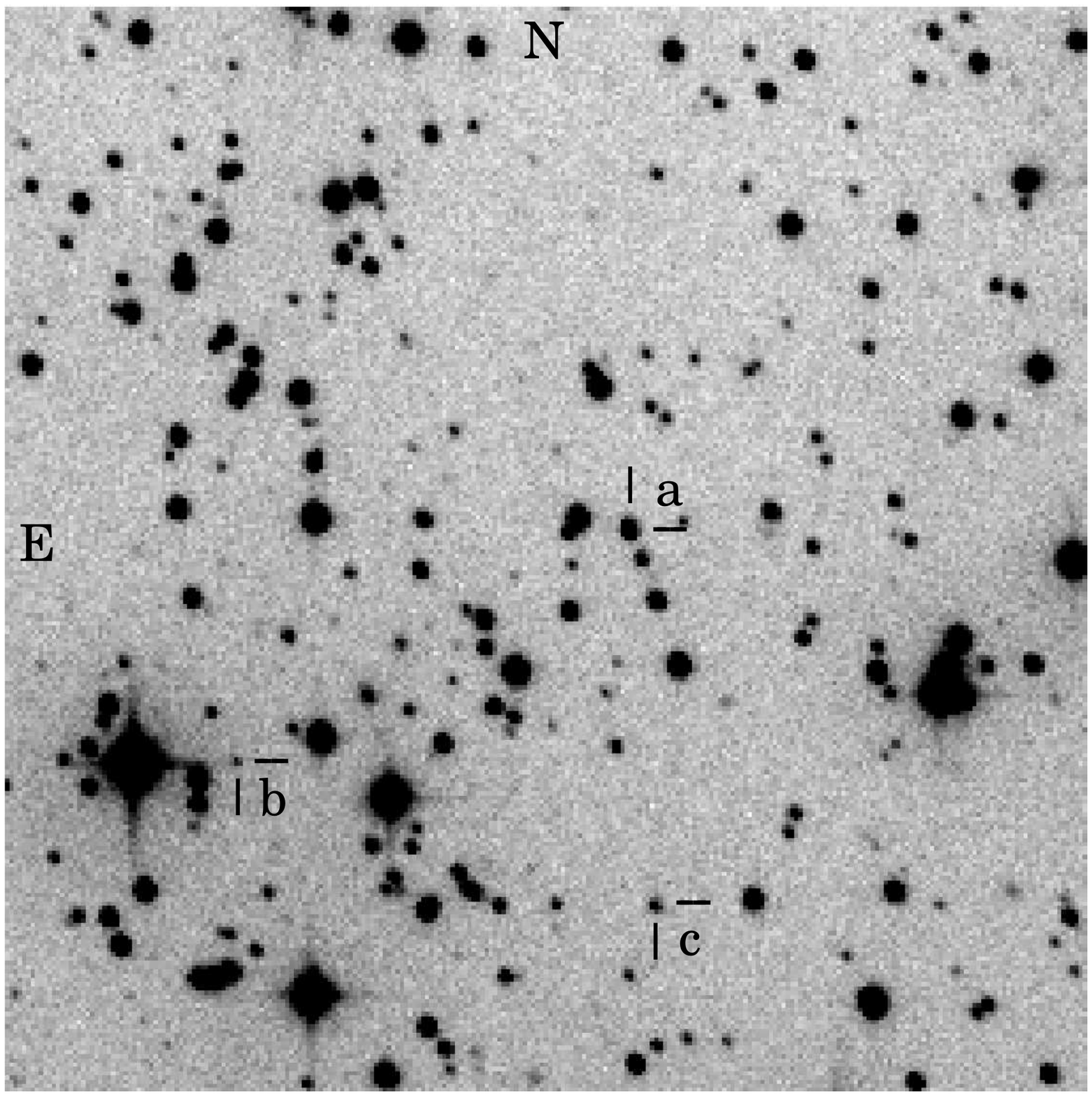}
}
\vspace{-0.7 truein}
\caption{Left: \xmm\ EPIC pn image (0.3--10~keV) from
2015 December 2 (ObsID 0790180101) showing the \fermi\
error circle of \fgl\ with a 95\% confidence radius of
$3.\!^{\prime}6$.
Right: Finding chart from an MDM 1.3m image taken through a BG38 filter.
The field is $4.\!^{\prime}3\times4.\!^{\prime}3$.
Labeled sources in both images are 
(a) the CV \rxs, (b) the MSP candidate \msp, and (c) the QSO \qso. 
}
\label{fig:finder}
\end{figure*}

\section{X-ray and Optical Observations}

The field of \fgl\ was observed twice with \xmm,
on 2015 October 20 (ObsID 0764420101, 53 ksec)
and 2015 December 2 (ObsID 0790180101, 77 ksec).
The EPIC pn detector was used in large window mode,
while the two MOS detectors were configured in small window mode,
all with the thin filter.  Time resolution is 48~ms for
the pn, and 0.3~s for the MOS.
In addition, the \xmm\ Optical Monitor (OM) obtained
10 contemporaneous exposures of 5000~s or 4160~s in the $V$-band
on 2015 October 20, and 15 contemporaneous 4400~s exposures in
the $B$-band on 2015 December 2.
Figure~\ref{fig:finder} (left) shows the
\fermi\ 95\% error circle superposed on the \xmm\ pn image
of 2015 December 2.
In addition to \rxs\ (source ``a''), there are two fainter
sources inside the error circle that we investigated,
labeled ``b'' and ``c''. The MOS small window only
included source ``a'',  but all three sources
have optical counterparts in the OM images.

We obtained another X-ray observation of \rxs\ using \chandra\
ACIS on 2016 July 7 for an exposure time of 30~ks (ObsID 17769).
The S3 CCD only was operated in continuous clocking (CC) mode
to avoid pileup of the bright source \rxs.  This gives time
resolution of 2.85~ms, but sacrifices one spatial dimension.
Only \rxs\ is bright enough to study in CC mode.

We obtained several hours of time-series optical photometry of the field
using the MDM Observatory 1.3m and 2.4m telescopes om Kitt Peak during
three observing runs in 2014, 2015, and 2016.  The detectors
used were either the $1024\times1024$ pixel thinned, back-illuminated
SITe CCD ``Templeton,''
or a thermoelectrically cooled Andor Ikon DU-937 N CCD camera.
The detectors, observing parameters, and differential photometry
techniques were the same as described in \citet{tho13} and \citet{hal15}.
The CCD readout was windowed and binned to reduce dead-time.
Exposure times were 10~s or 20~s, with 3~s dead-time using
Templeton, and 5~s exposures with 12~ms dead-time using the Andor.
A log of the observations is given in Table~\ref{tab:optlog}.
The 1.3m Templeton images (Figure~\ref{fig:finder}, right)
covered all three X-ray sources, while the Andor and 2.4m Templeton
images only included \rxs.
Optical positions for the three sources, measured from
a 1.3m image using USNO B1.0 catalog stars for the
astrometric solution, are given in Table~\ref{tab:positions}.

We obtained 59 spectra of \rxs\ during two observing runs,
in 2016 January and February, using the 2.4m telescope 
with the modular spectrograph
and a thinned, back-illuminated $2048\times2048$ pixel SITe CCD detector.
Our spectra covered
4210--7500\,\AA, with 2.0~\AA\ pixel$^{-1}$ and a FWHM resolution
of $\approx3.5$\,\AA .  The observing, reduction, and analysis protocols
were practically identical to those used in \citet{tho13} and 
\citet{hal15}.  A log of the spectroscopic observations
is given in Table~\ref{tab:radial}.
Finally, we obtained one identification spectrum of source ``c'' using
the Ohio State Multi-Object Spectrograph (OSMOS) on the 2.4m.

\begin{figure}
\centerline{
\includegraphics[width=1.1\linewidth,clip=true]{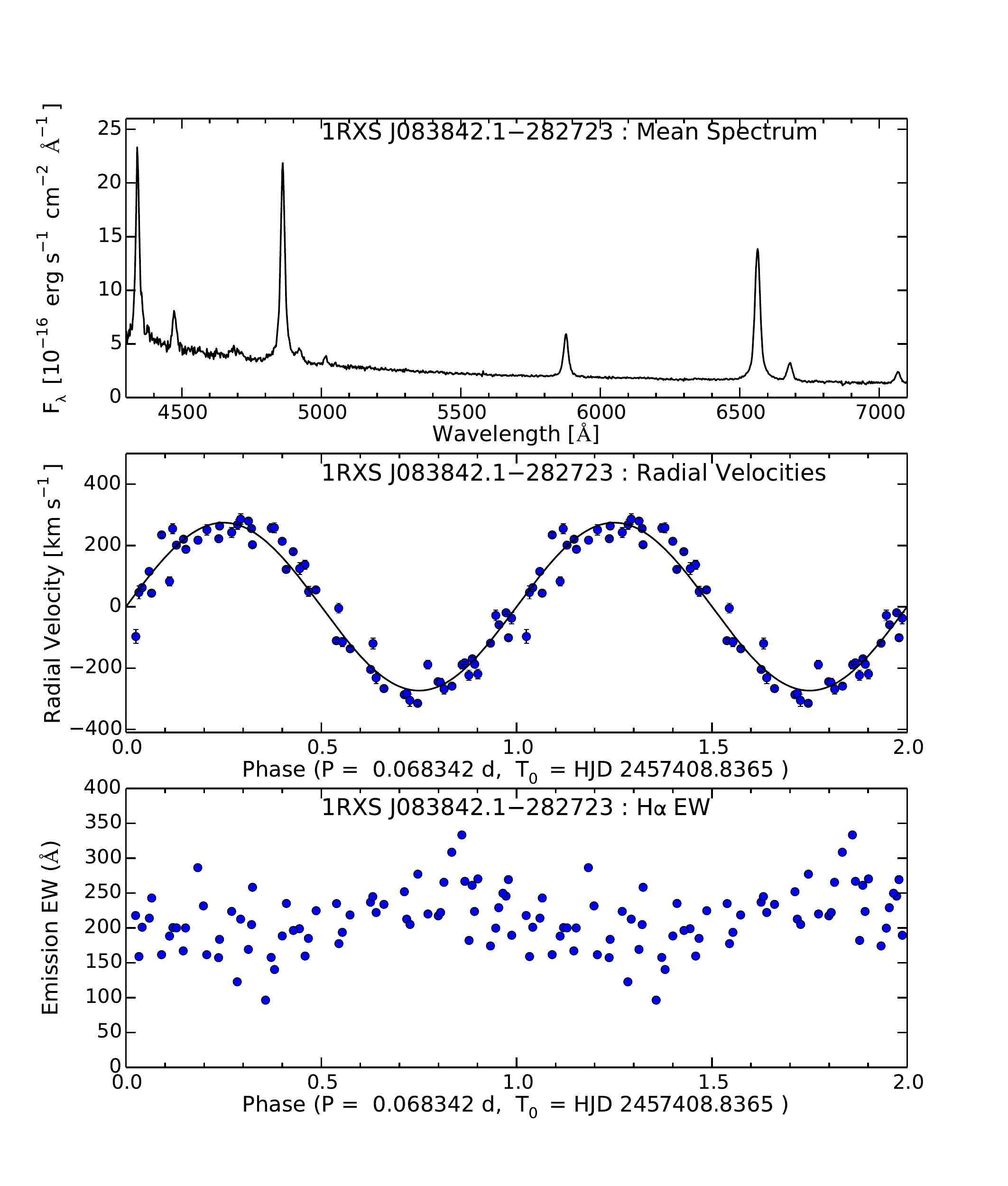}
}
\vspace{-0.25 truein}
\caption{Top: Mean spectrum of \rxs.  Middle: Radial velocities
of the H$\alpha$ emission line, folded on the best-fitting period, together
with a sinusoidal fit.  The error bars shown are based on signal-to-noise
and do not include systematic effects.  
The data are repeated for a second cycle for continuity. 
Bottom: Equivalent width of the H$\alpha$ emission line.
}
\label{fig:specfold}
\end{figure}

\begin{figure}
\centerline{
\includegraphics[width=1.15\linewidth,clip=true]{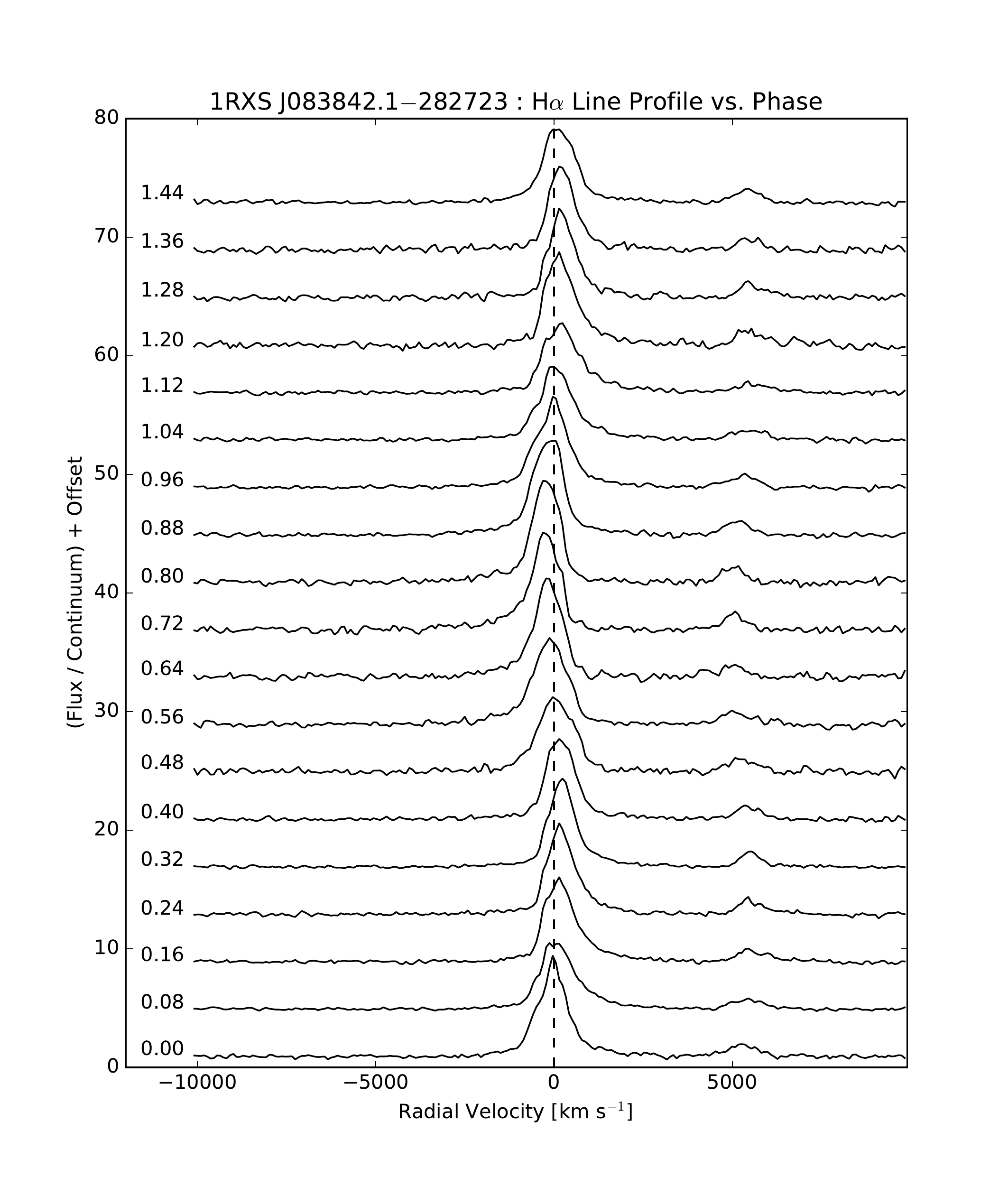}
}
\vspace{-0.25 truein}
\caption{
Rectified spectra of \rxs\ near H$\alpha$, as a function of orbital phase.  Each trace
is a weighted average of spectra taken near the phase indicated.  
Successive traces have been shifted upward by four times the continuum.
}
\label{fig:halphaprofiles}
\end{figure}

\begin{deluxetable}{ccccc}
\tablewidth{0pt}
\tablecaption{H$\alpha$ Spectroscopy of \rxs}
\tablehead{\colhead{Date} &
\colhead{Exposure} &
\colhead{$v_r$} & 
\colhead{$\sigma$} &
\colhead{H$\alpha$ EW} \\
\colhead{(BJD)\tablenotemark{a}} &
\colhead{(s)} &
\colhead{(km~s$^{-1}$)} & 
\colhead{(km~s$^{-1}$)} &
\colhead{(\AA)}
}
\startdata
2457402.8178  & 900  &   $ -119$ & $   6$ & 174 \\
2457402.8286  & 900  &   $  234$ & $   8$ & 162 \\
2457403.8843  & 480  &   $ -110$ & $   6$ & 235 \\
2457403.8903  & 480  &   $ -204$ & $   6$ & 237 \\
2457403.8962  & 480  &   $ -287$ & $   7$ & 252 \\
2457403.9021  & 480  &   $ -244$ & $   6$ & 217 \\
2457403.9081  & 480  &   $ -170$ & $   6$ & 261 \\
2457403.9140  & 480  &   $  -19$ & $   6$ & 246 \\
2457403.9199  & 480  &   $  115$ & $   8$ & 214 \\
2457403.9259  & 480  &   $  220$ & $   9$ & 167 \\
2457405.7628  & 480  &   $  -97$ & $  23$ & 218 \\
2457405.7687  & 480  &   $   83$ & $  15$ & 188 \\
2457405.7806  & 480  &   $  268$ & $  16$ & 123 \\
2457405.7865  & 480  &   $  257$ & $  14$ & 158 \\
2457405.7924  & 480  &   $  137$ & $  15$ & 160 \\
2457405.7983  & 480  &   $   -5$ & $  16$ & 177 \\
2457405.8043  & 480  &   $ -120$ & $  18$ & 245 \\
2457405.8102  & 480  &   $ -282$ & $  14$ & 213 \\
2457405.8161  & 480  &   $ -247$ & $  13$ & 222 \\
2457405.8221  & 480  &   $ -187$ & $  12$ & 224 \\
2457405.8280  & 480  &   $ -101$ & $   9$ & 269 \\
2457405.8339  & 480  &   $   44$ & $  10$ & 243 \\
2457405.8399  & 480  &   $  188$ & $   9$ & 200 \\
2457405.8458  & 480  &   $  263$ & $   9$ & 184 \\
2457405.9192  & 480  &   $  279$ & $   8$ & 169 \\
2457405.9251  & 480  &   $  214$ & $   7$ & 188 \\
2457405.9310  & 480  &   $   55$ & $   7$ & 225 \\
2457405.9370  & 480  &   $ -137$ & $   7$ & 219 \\
2457405.9429  & 480  &   $ -267$ & $   8$ & 234 \\
2457405.9488  & 480  &   $ -315$ & $   7$ & 277 \\
2457405.9548  & 480  &   $ -259$ & $   8$ & 308 \\
2457407.7556  & 480  &   $  217$ & $  12$ & 286 \\
2457407.7615  & 480  &   $  243$ & $  16$ & 224 \\
2457407.7734  & 480  &   $  125$ & $  20$ & 199 \\
2457407.8030  & 480  &   $ -223$ & $  17$ & 182 \\
2457408.7526  & 480  &   $ -188$ & $  14$ & 220 \\
2457408.7586  & 480  &   $ -189$ & $  13$ & 333 \\
2457408.7645  & 480  &   $  -28$ & $  17$ & 200 \\
2457408.7704  & 480  &   $   47$ & $  21$ & 159 \\
2457408.7763  & 480  &   $  255$ & $  16$ & 200 \\
2457408.7823  & 480  &   $  251$ & $  17$ & 162 \\
2457408.7882  & 480  &   $  286$ & $  17$ & 213 \\
2457408.7941  & 480  &   $  258$ & $  17$ & 140 \\
2457408.8001  & 480  &   $   50$ & $  16$ & 185 \\
2457408.8060  & 480  &   $ -115$ & $  15$ & 194 \\
2457408.8119  & 480  &   $ -232$ & $  19$ & 222 \\
2457408.8178  & 480  &   $ -305$ & $  20$ & 205 \\
2457408.8238  & 480  &   $ -268$ & $  16$ & 265 \\
2457408.8297  & 480  &   $ -219$ & $  16$ & 270 \\
2457408.8356  & 480  &   $  -37$ & $  19$ & 189 \\
2457430.8336  & 480  &   $ -183$ & $   4$ & 267 \\
2457430.8396  & 480  &   $  -59$ & $   5$ & 229 \\
2457430.8455  & 480  &   $   62$ & $   5$ & 201 \\
2457430.8515  & 480  &   $  201$ & $   5$ & 200 \\
2457438.7866  & 480  &   $  222$ & $   6$ & 157 \\
2457438.7925  & 480  &   $  202$ & $   8$ & 258 \\
2457438.7984  & 480  &   $  122$ & $   6$ & 235 \\
2457440.8426  & 600  &   $  255$ & $   6$ & 205 \\
2457440.8499  & 600  &   $  180$ & $   6$ & 196
\enddata
\tablenotetext{a}{Barycentric Julian day of mid-integration in the UTC system.}
\label{tab:radial}
\end{deluxetable}

\begin{figure}
\centerline{
\includegraphics[angle=0,width=1.2\linewidth,clip=]{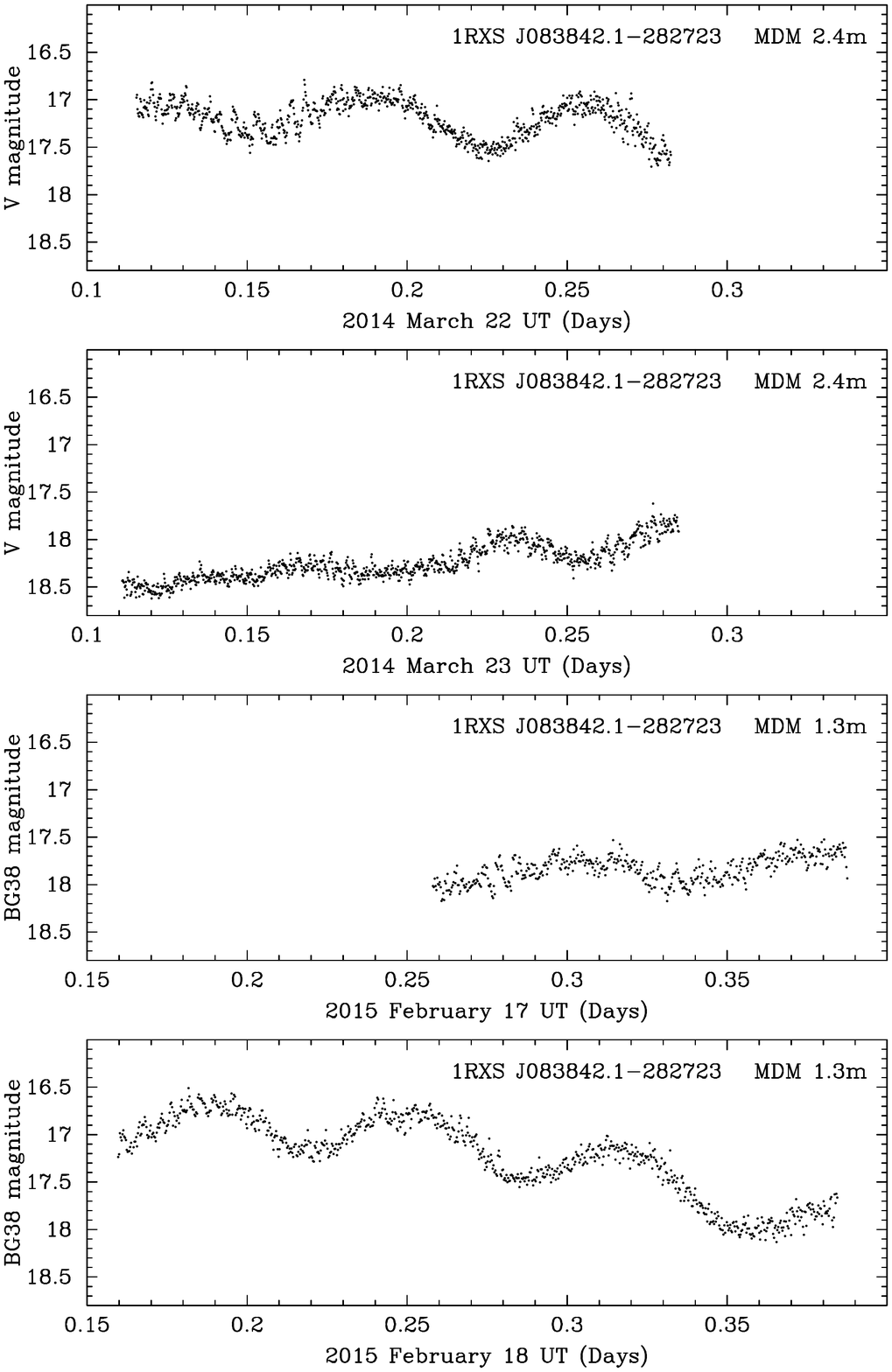}
}
\hspace{-0.15in}
\centerline{
\includegraphics[angle=0,width=1.1\linewidth,clip=]{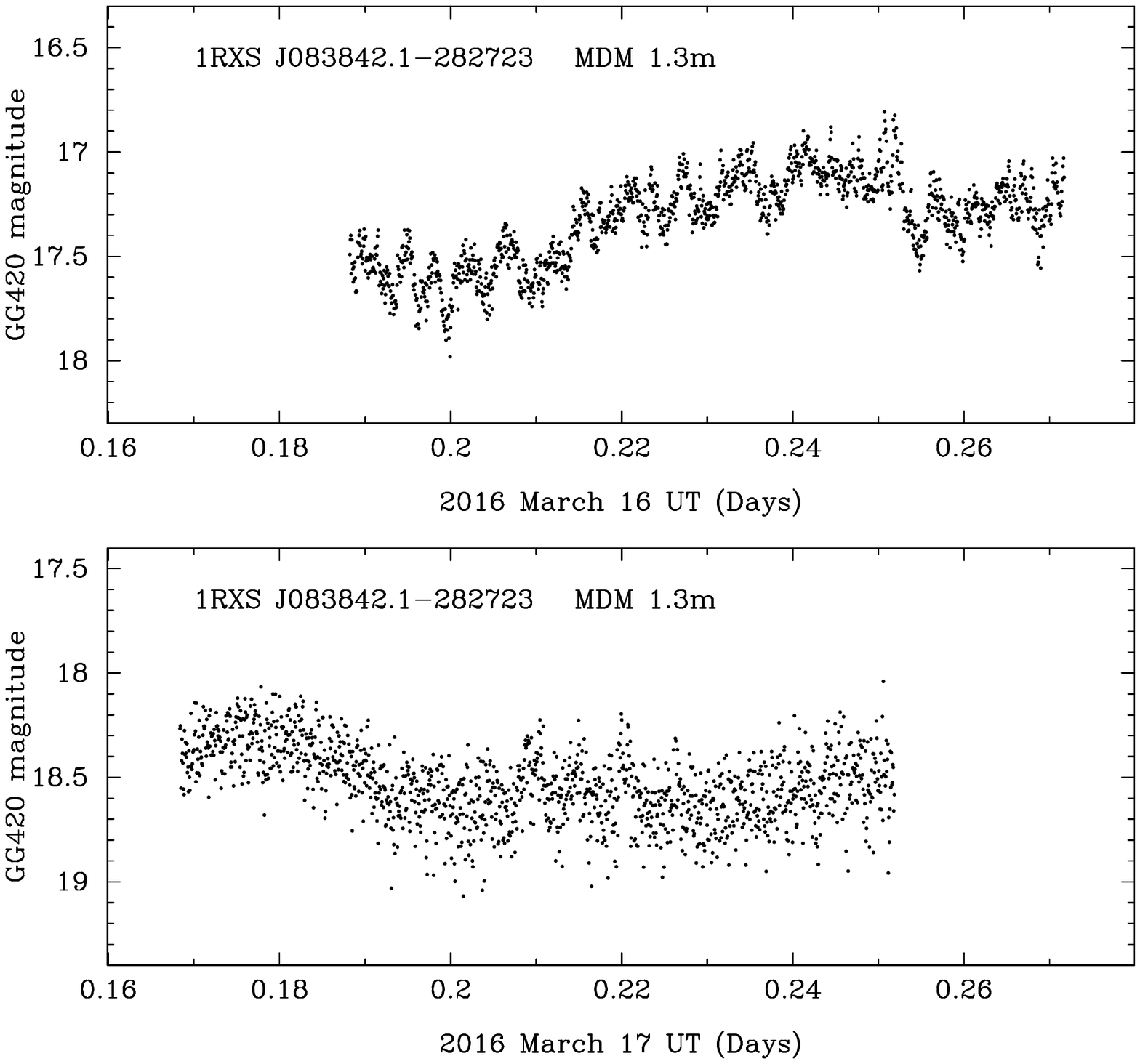}
}
\vspace{-1.6in}
\caption{MDM light curves of \rxs; the log of observations is
given in Table~\ref{tab:optlog}.  Times are barycentric dynamical
time (TDB).
}
\label{fig:optphot}
\end{figure}

\section{\rxs}

\subsection{Optical Spectroscopy}

Our mean spectrum of \rxs\ (Figure~\ref{fig:specfold}, upper panel) 
appears similar to that published by \citet{mas13}, with 
strong, single-peaked Balmer and \ion{He}{1} lines on a blue continuum.
\ion{He}{2}~$\lambda4686$ is just detected with an emission equivalent 
width (EW) of $\sim8$\,\AA, but is much less prominent than
H$\beta$ (EW $\approx125$\,\AA).
The spectrum implies a synthetic $V\sim18.0$, but this is 
not expected to be precise because of occasional clouds and 
poor seeing.  

We measured radial velocities of H$\alpha$ using a convolution
function tuned to be sensitive to the steep sides of the line
profile, which are about 1500~km~s$^{-1}$ apart.  Table~\ref{tab:radial}
lists the resulting radial velocities and uncertainties.  To find the 
period we constructed a dense grid of trial frequencies and 
fitted the velocities with least-squares sinusoids at each
frequency.  This yielded a period near 98.4 minutes,
with no ambiguity in the daily or month-to-month cycle count. 
A least-squares best fit of the form
$v_r(t) = \gamma + K \sin[2 \pi (t - T_0)/P]$ has
\begin{eqnarray}
T_0 &=& {\rm BJD}\ 2457408.8365 \pm 0.0006 \\
P &=& 0.068342(3) \ {\rm d} \\
K &=& 274 \pm 16\ {\rm km\ s^{-1}} \\
\gamma &=& 49 \pm 11\ {\rm km\ s^{-1}}
\end{eqnarray}
with an RMS scatter of 36~km~s$^{-1}$ for the 59 data
points. The precise period is \optper.
The middle panel of Figure~\ref{fig:specfold} shows
the sine fit superposed on the radial velocities, and
Figure~\ref{fig:halphaprofiles} shows the H$\alpha$
line profiles as function of orbital phase.

The lower panel of Figure~\ref{fig:specfold}
shows the equivalent width (EW) of the H$\alpha$ emission
line as a function of spectroscopic phase.  A periodogram
of the EW has a peak at exactly the spectroscopic period.
Although noisy, the EW varies sinusoidally with a phase lag
of $\approx0.5$ cycles with respect to the radial velocity curve.
As will be discussed, this is a clue to the accretion geometry.

\subsection{Time-series Optical Photometry}

The MDM light curves of \rxs\ are shown in Figure~\ref{fig:optphot}.
Magnitudes were calibrated to stars in the UCAC4 \citep{zac13}.
BG38 and GG420 are broadband filters chosen for their higher throughput,
but they do not have standard calibrations.  We approximated
their magnitudes using $V$ and $R$, respectively,
from stars in the UCAC4.

The individual time series all display a broad oscillation with a period
of $\sim0.07$~d, similar to the spectroscopic period.
Although some of the light curves in
Figure~\ref{fig:optphot} show a hint of a faster oscillation,
a power-spectrum analysis does not reveal any shorter coherent period.
In addition, in all three years there is a large change of $\approx1.5$
magnitudes between adjacent nights.  When the star is bright,
the $\sim0.07$~d oscillation has large amplitude; when it is faint
the relative amplitude is smaller.  This behavior matches the X-ray 
(presented in the next section) very well, with the night-to-night changes
explained as modulation of the accretion rate on the
beat period between the spin and the orbit of a stream-fed system.

\begin{deluxetable*}{clccccc}[th]
\tabletypesize{\small}
\tablewidth{0pt}
\tablecaption{X-ray Spectral Fits}
\tablehead{
\colhead{Label} & \colhead{Source} & \colhead{$N_{\rm H}$ (cm$^{-2}$)} &
\colhead{$kT_{\rm br}$ (keV)} & \colhead{$\Gamma$} &
\colhead{$F_x$ (0.3--10 keV)\tablenotemark{a}} &
\colhead{$\chi^2_{\nu}\ (\rm dof)$}
}
\startdata
\cutinhead{\chandra\ ACIS-S, ObsID 17769, 2016 July 7}
a  & \rxs\  &  $5.6_{-1.4}^{+1.5}\times10^{20}$ &  $11.2_{-1.1}^{+1.3}$  & ... & $(9.43\pm0.14)\times10^{-12}$ & 1.25 (368) \\
\cutinhead{\xmm\ EPIC-pn, ObsID 0790180101, 2015 December 2}
a  & \rxs\  &  $1.03_{-0.20}^{+0.21}\times10^{20}$ &  $11.7\pm0.5$  & ... & $(6.61\pm0.06)\times10^{-12}$ & 1.48 (883) \\
b  & \msp\  &  $1.19_{-0.33}^{+0.37}\times10^{21}$ & ... & $1.5\pm0.1$ & $(1.73\pm0.11)\times10^{-13}$ & 0.96 (109)\\
b  & \msp\  &  $(7.1\pm2.1)\times10^{20}$ & $14.4_{-5.2}^{+11.1}$ & ... & $(1.63\pm0.13)\times10^{-13}$ & 0.99 (109)\\
c  & \qso\  &  $1.26_{-0.59}^{+0.69}\times10^{21}$ & ... & $2.2\pm0.3$ & $7.99_{-0.08}^{+0.11}\times10^{-14}$ & 0.69 (44) 
\enddata
\tablecomments{Uncertainties are 90\% confidence.}
\tablenotetext{a}{Unabsorbed flux in units of erg cm$^{-2}$ s$^{-1}$.}
\label{tab:spectra}
\end{deluxetable*}

\begin{figure}
\vspace{-0.25in}
\centerline{
\hspace{0.1in}
\includegraphics[angle=0,width=1.13\linewidth,clip=]{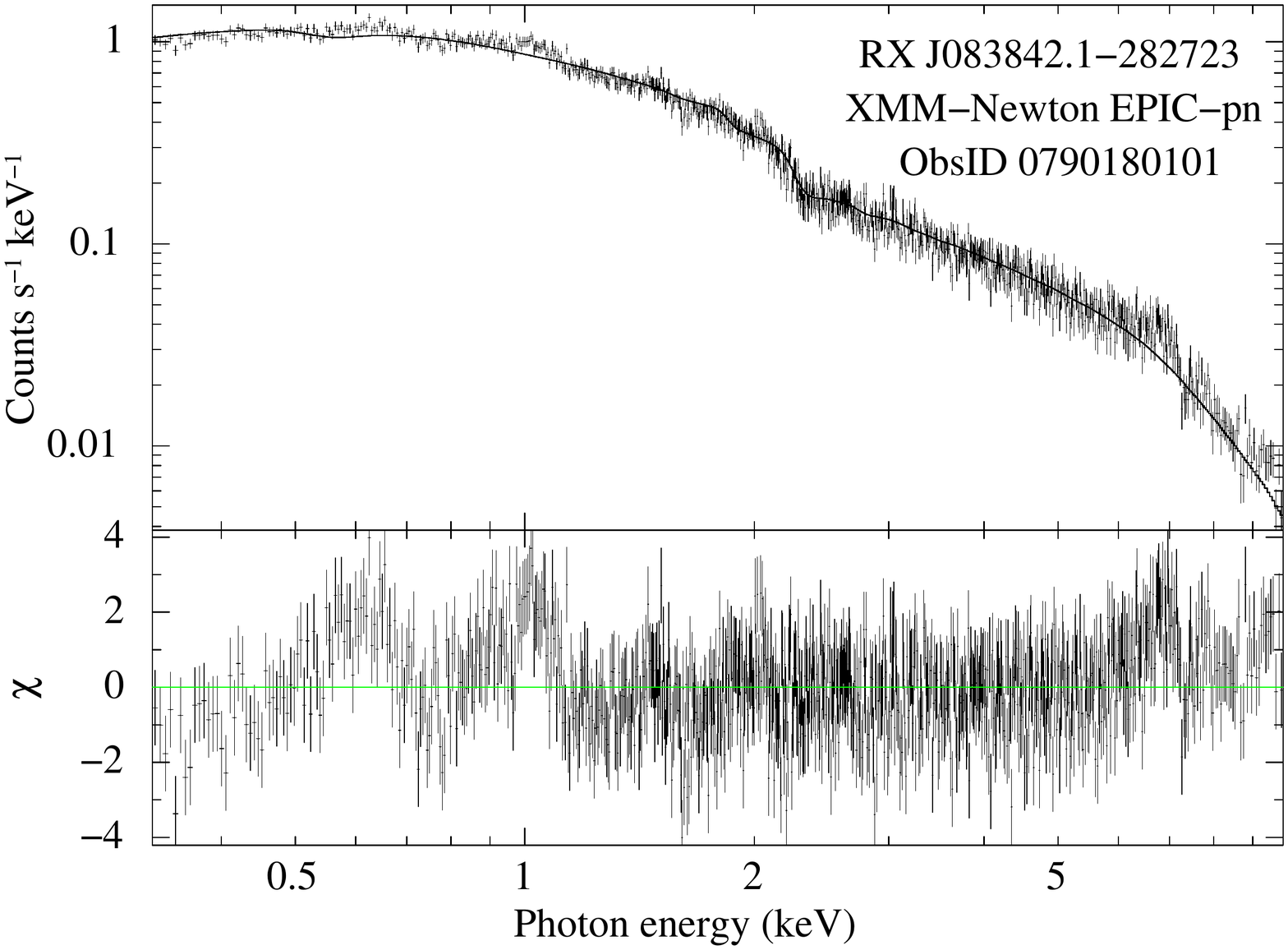}
}
\vspace{-0.4 truein}
\hspace{0.1in}
\centerline{
\includegraphics[angle=0,width=1.13\linewidth,clip=]{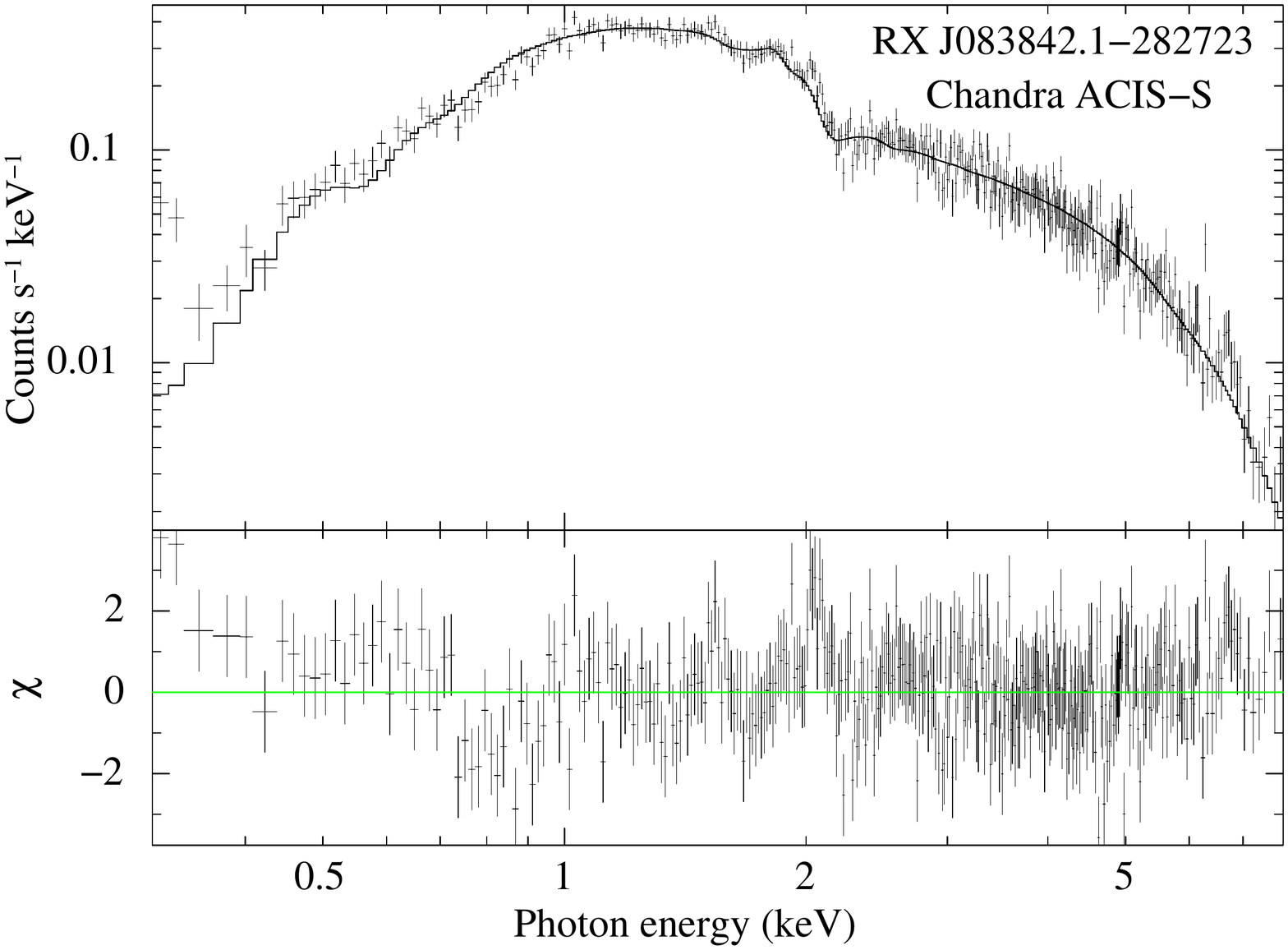}
}
\vspace{-0.1 truein}
\caption{X-ray spectra of \rxs\ fitted to a thermal bremsstrahlung
model.  Parameters of the fit are listed in Table~\ref{tab:spectra}.
Residuals correspond to Fe~K$\alpha$ 6.7, 6.9 keV, as well as
Fe L-shell (1 keV)  and O K-shell (0.6 keV) emission lines from the
hot plasma. 
}
\label{fig:rxs_spec}
\end{figure}

\begin{figure*}
\vspace{-0.6in}
\centerline{
\includegraphics[angle=0,width=1.05\linewidth,clip=]{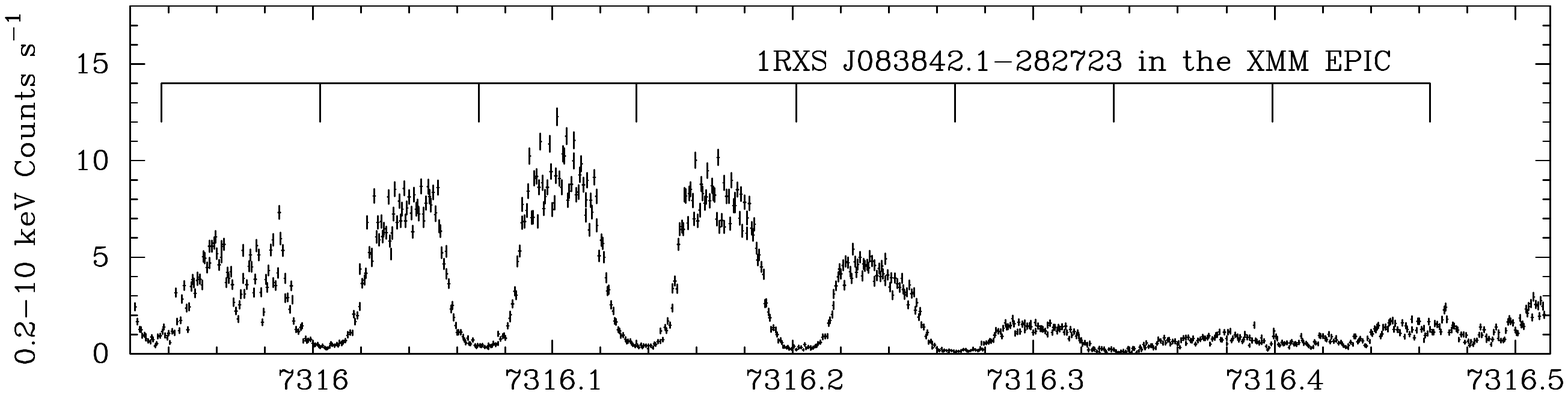}
}
\vspace{-4. truein}
\centerline{
\includegraphics[angle=0,width=1.05\linewidth,clip=]{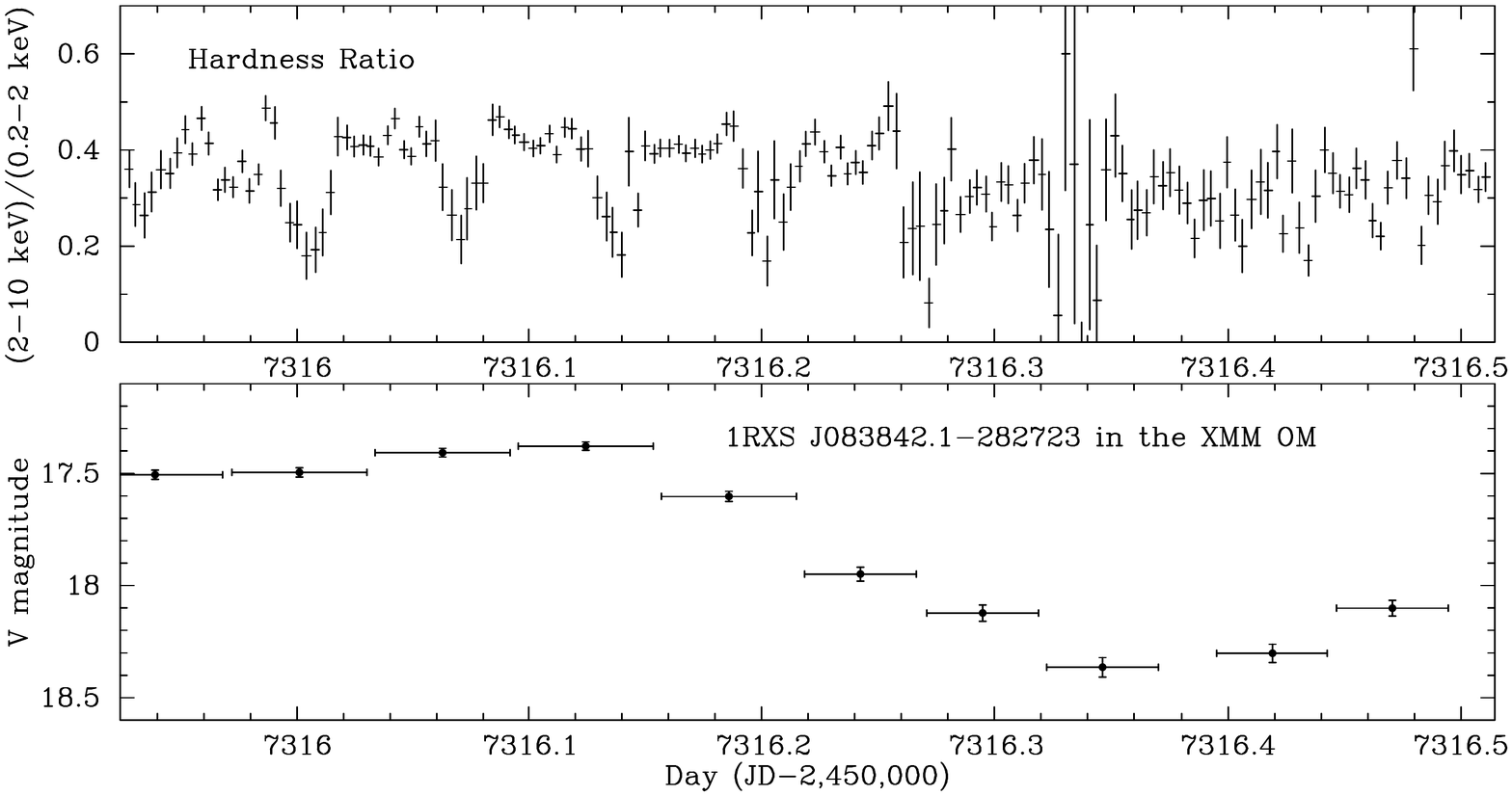}
}
\vspace{-1.8 truein}
\caption{Background-subtracted \xmm\ light curve of \rxs\ on 2015 October 20
(ObsID 0764420101).   Top: Combined EPIC pn and MOS broad-band
(0.2--10~keV) light curve in 60~s bins.
The 95 minute period is indicated by tick marks, which reveal the
$\approx120^{\circ}$ phase jump after minimum light of the beat cycle.
Middle: Hardness ratio of counts in the (2--10 keV)/(0.2--2 keV) bands,
in 300~s bins.
Bottom: \xmm\ OM magnitudes from 5000~s or 4160~s exposures in the $V$-band.
}
\label{fig:cv_light_curve_2}
\end{figure*}

\begin{figure*}
\centerline{
\includegraphics[angle=0,width=1.05\linewidth,clip=]{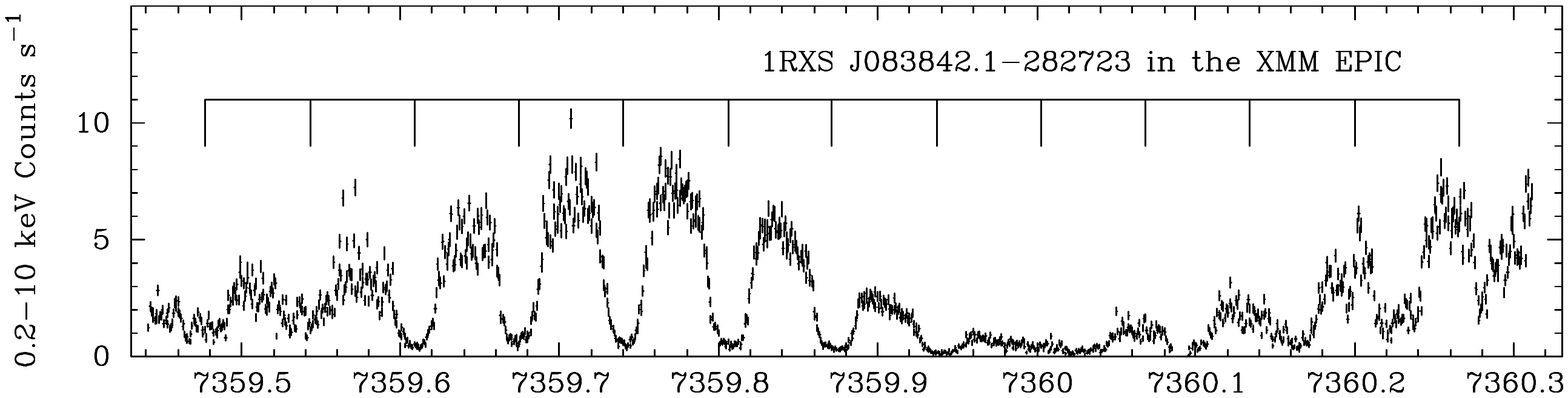}
}
\vspace{-4.truein}
\centerline{
\includegraphics[angle=0,width=1.05\linewidth,clip=]{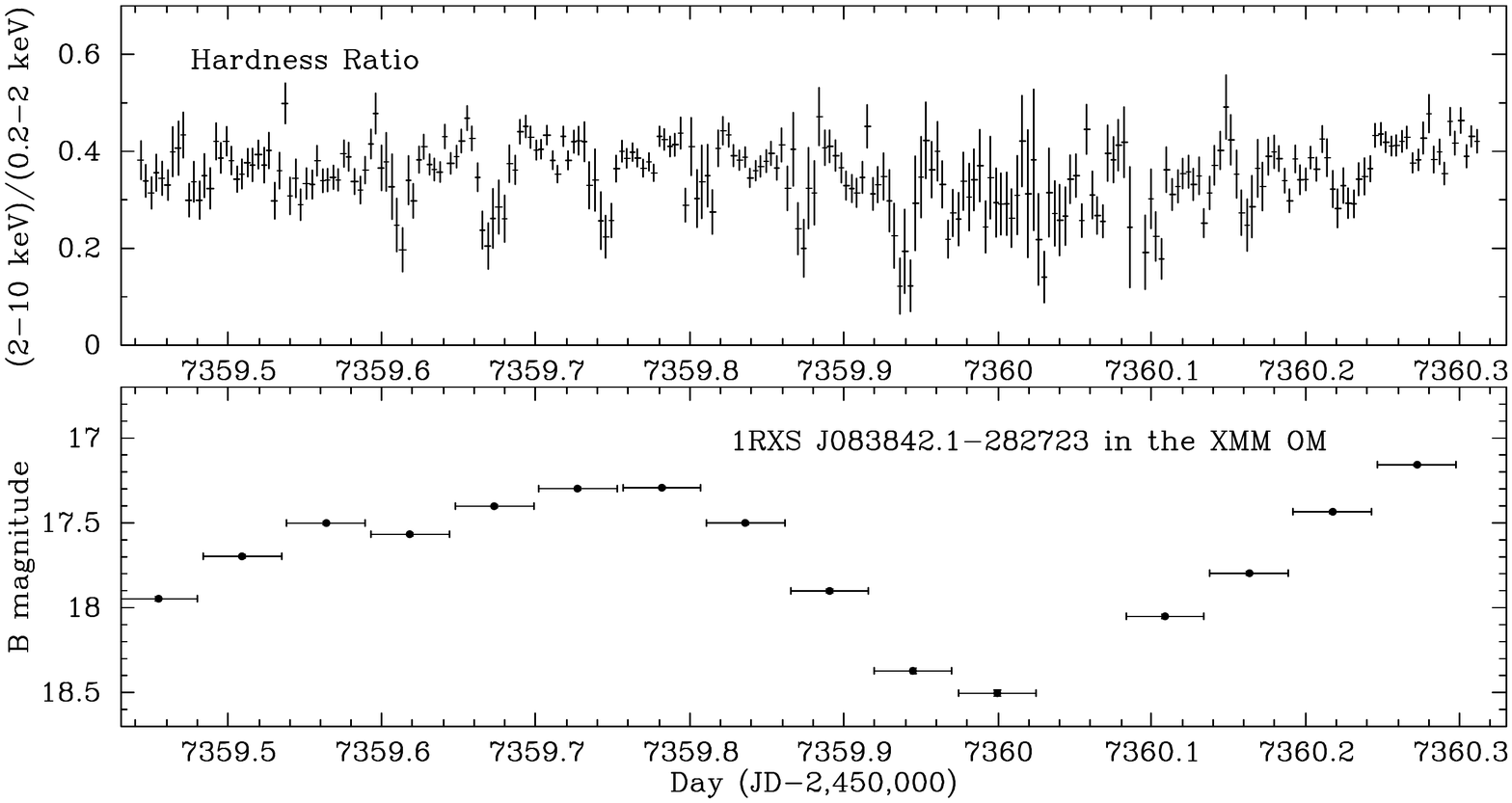}
}
\vspace{-1.8 truein}
\caption{Background-subtracted \xmm\ light curve of \rxs\ on 2015 December 2
(ObsID 0790180101). 
Top: Combined EPIC pn and MOS broad-band (0.2--10~keV) light curve in 60~s bins.
There is a short gap in the data at day 7360.09.
The 94.6 minute period is indicated by tick marks, which reveal
the $\approx120^{\circ}$ phase jump after minimum light of the beat cycle.
Middle: Hardness ratio of counts in the (2--10 keV)/(0.2--2 keV) bands,
in 300~s bins.
Bottom: \xmm\ OM magnitudes from 4400~s exposures in the $B$-band.
There is a gap in the data at day 7360.05.
}
\label{fig:cv_light_curve_1}
\end{figure*}

\begin{figure*}
\vspace{-0.3in}
\centerline{
\hspace{0.2in}
\includegraphics[angle=0,width=0.58\linewidth,clip=]{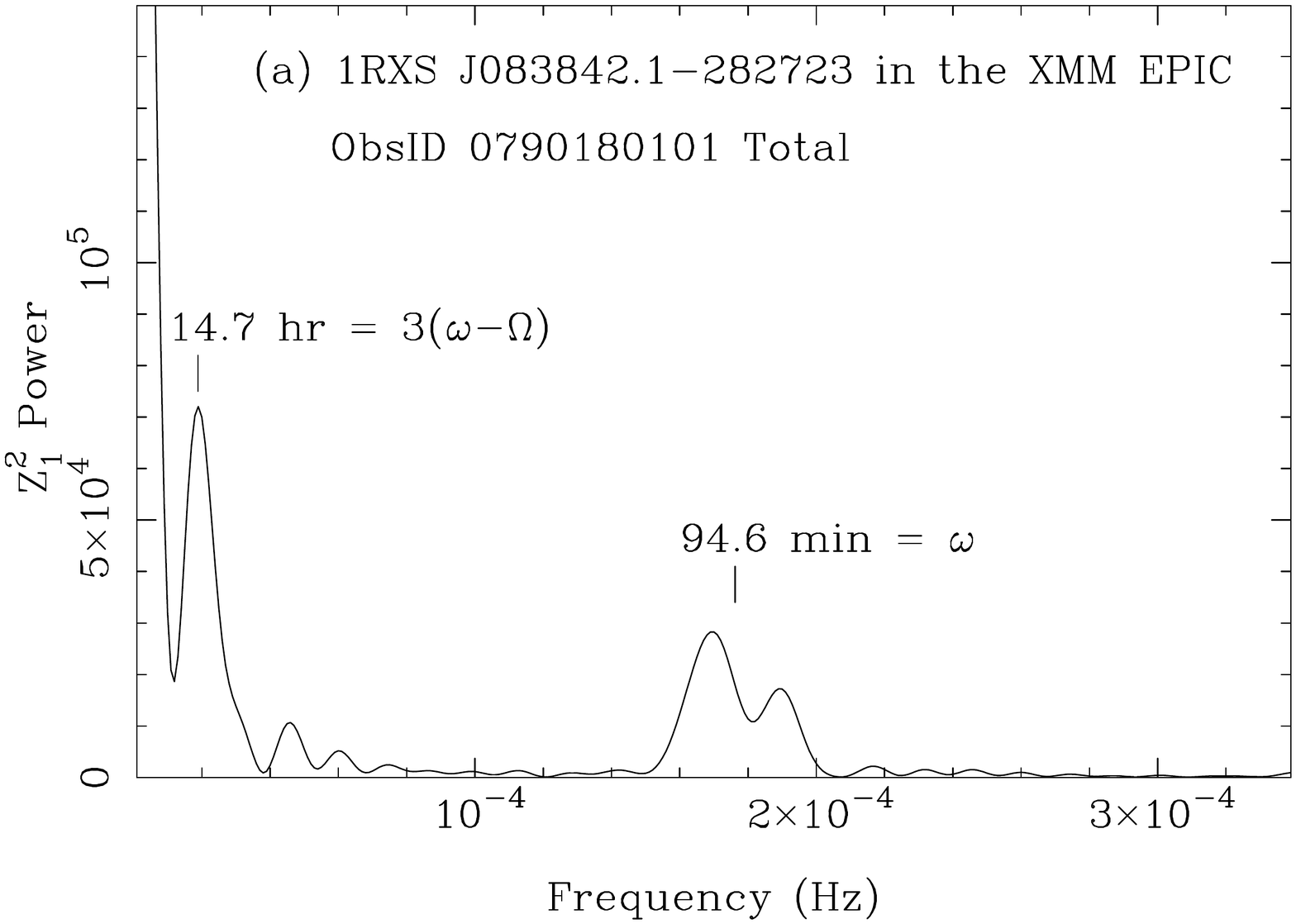}
\hspace{-0.65 truein}
\includegraphics[angle=0,width=0.58\linewidth,clip=]{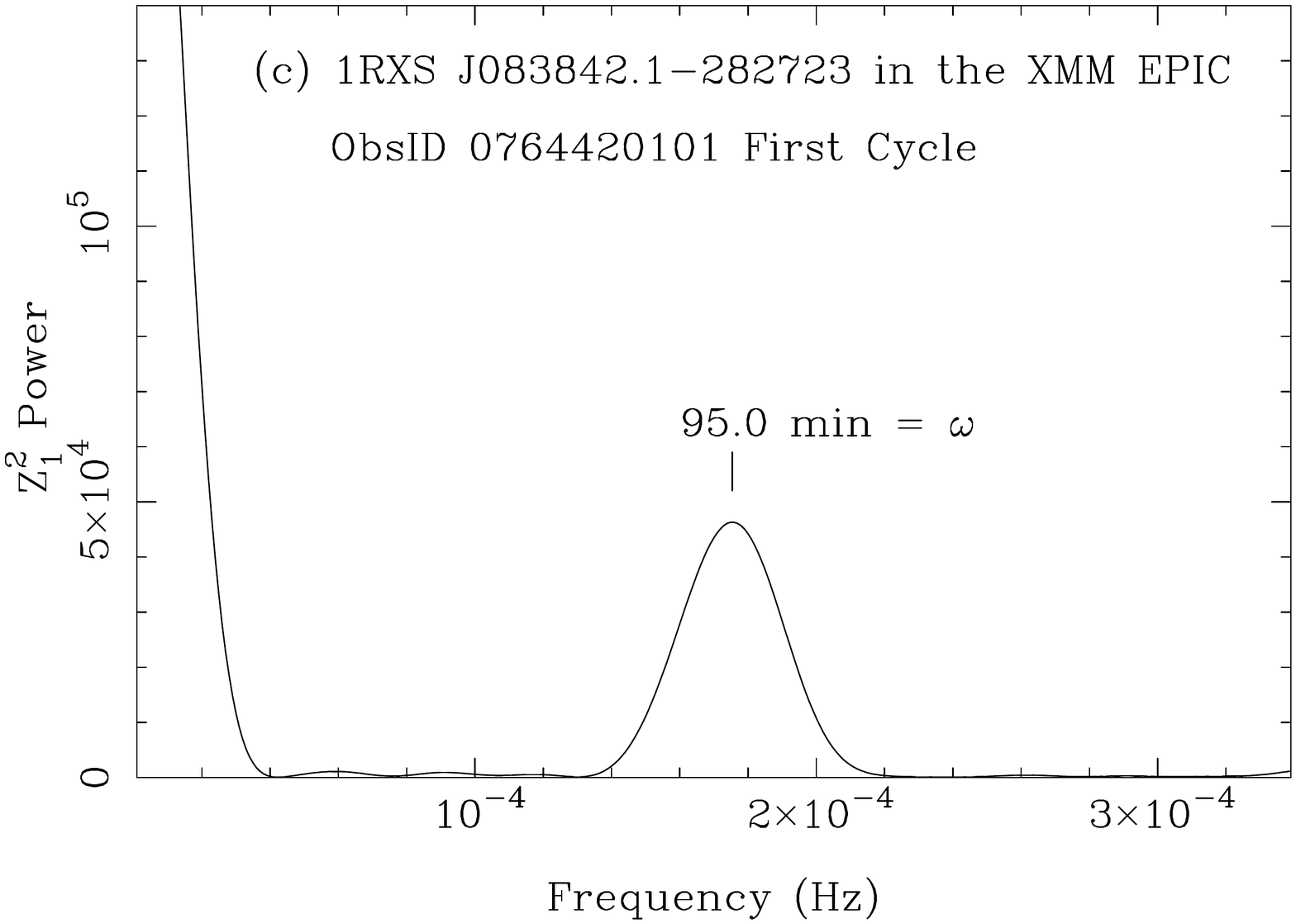}
}
\vspace{-0.6 truein}
\centerline{
\hspace{0.2in}
\includegraphics[angle=0,width=0.58\linewidth,clip=]{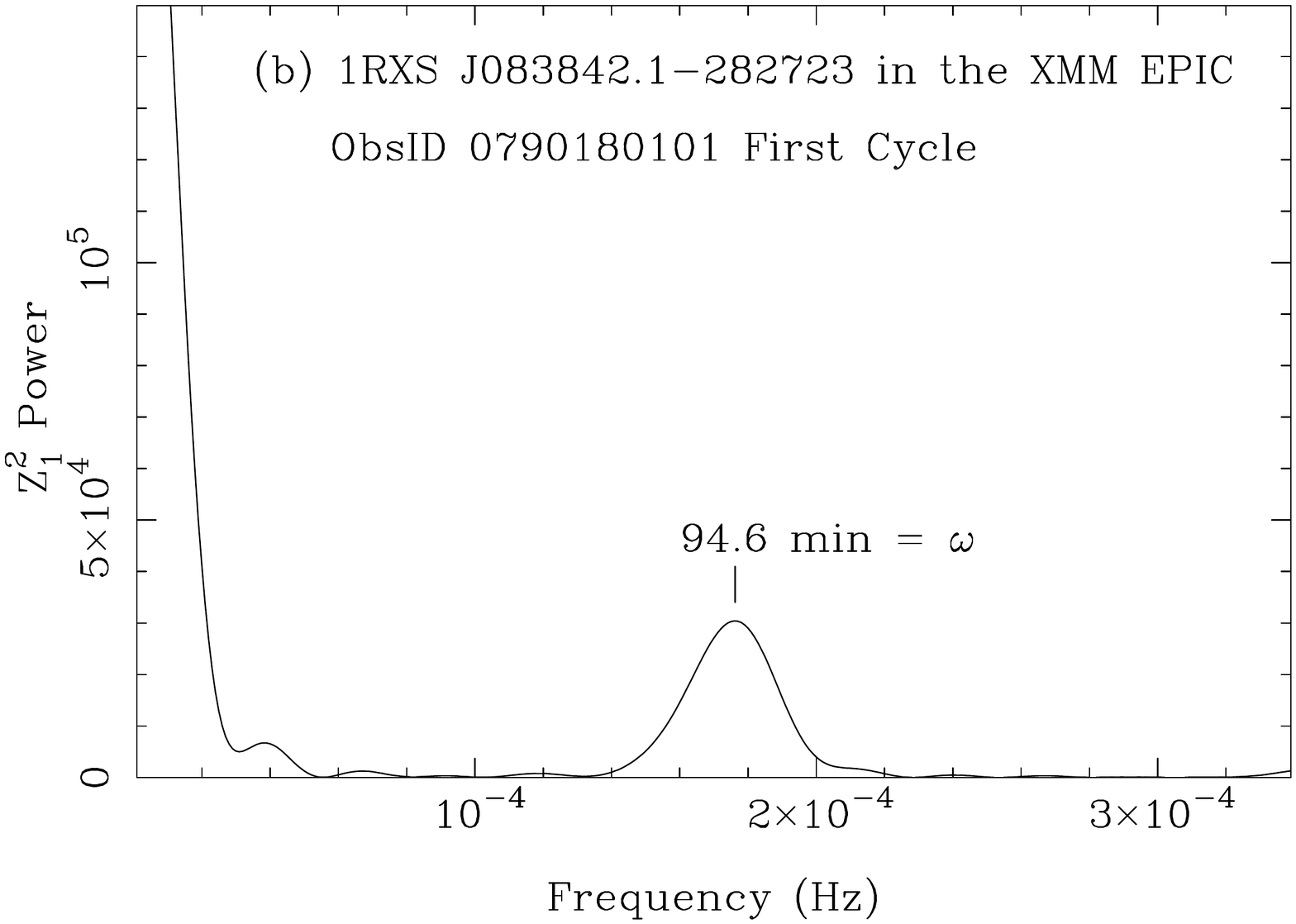}
\hspace{-0.65 truein}
\includegraphics[angle=0,width=0.58\linewidth,clip=]{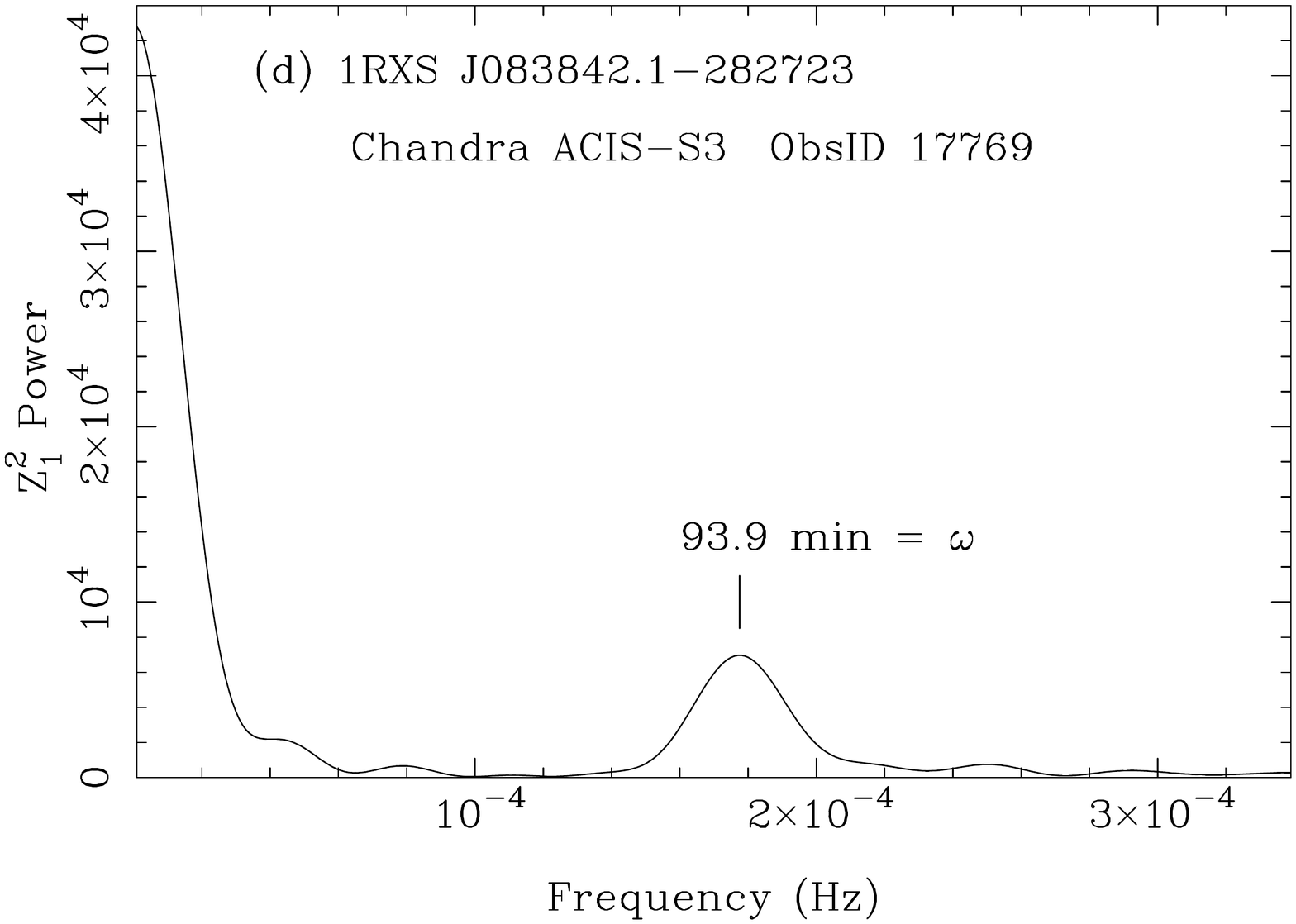}
}
\vspace{-0.15in}
\caption{Power spectra of the X-ray light curves of \rxs.
(a) From the \xmm\ observation of Figure~\ref{fig:cv_light_curve_1},
the 94.6~minute spin period is split into peaks at 98.3~minutes and
88.0~minutes by the phase jump after the minimum of the beat cycle
at day 7360.  The 14.7~hr period is inferred to be one-third 
of the beat cycle because the phase jump indicates that there is
pole switching spanning $\approx120^{\circ}$.
The spin frequency is $\omega$ and the inferred orbital frequency
is $\Omega$.
(b) When the power spectrum is restricted to the first beat cycle,
before day 7360 in Figure~\ref{fig:cv_light_curve_1},
a single peak at the true period of 94.6~minutes is recovered.
(c) The power spectrum of the \xmm\ observation of
Figure~\ref{fig:cv_light_curve_2}, restricted to the first beat cycle
(before day 7316.4).  The 95 minute period is consistent with the
other X-ray observations.
(d) The power spectrum of the \chandra\ observation of
Figure~\ref{fig:chandra_light_curve}, which does not span
a minimum of the beat cycle.  The 93.9 minute period is consistent
with the other X-ray observations.  We adopt $94.8\pm0.4$ minutes
as the spin period.}
\label{fig:xmm_power}
\end{figure*}

\subsection{\xmm\ Analysis}

We fitted the pn X-ray spectrum from the longer \xmm\ observation with a
thermal bremsstrahlung model in XSPEC.  The temperature is 11.7~keV
(Figure~\ref{fig:rxs_spec} and Table~\ref{tab:spectra}),
and residuals corresponding to Fe~K$\alpha$ 6.7, 6.9 keV,
as well as Fe L-shell and O K-shell lines, are evident.
A {\tt mekal} hot plasma
model gives a similar temperature.  There is no evidence of a soft
blackbody from the white dwarf (WD) surface that is sometimes
but not always seen in polars.

X-ray photon arrival times were transformed to Barycentric Dynamical Time
and extracted from a $20^{\prime\prime}$ radius around \rxs.
Figures~\ref{fig:cv_light_curve_2} and \ref{fig:cv_light_curve_1}
show the 0.2--10~keV, background subtracted,
combined light curve from the pn and MOS in 60~s bins,
as well as magnitudes from the OM.
The X-ray light curve shows a factor of
10 variation with a period of
$\approx0.07$~d, as well as a broader modulation on a timescale
of $\sim0.6$~d.  This resembles a classic beating
between two closely spaced frequencies.
A $Z_1^2$ periodogram of the longer (December 2) observation
(Figure~\ref{fig:xmm_power}a) shows three
peaks, at 88.0 minutes, 98.3 minutes, and 14.7~hr, where
the latter is consistent with the beat between the two
shorter periods.  However, a periodogram of the shorter
observation (October 20; Figure~\ref{fig:xmm_power}c) has only
a single peak at a period of 95 minutes, which falls in between the pair
at 88.0 minutes and 98.3 minutes.

Inspection of the light curves reveals what is responsible
for the difference in the power spectra.
At the minimum of the 14.7~hr cycle,
the phase of the shorter period jumps by $\approx120^{\circ}$.
This is especially clear in Figure~\ref{fig:cv_light_curve_1},
where the periodic tick marks switch from
marking flux minima before day 7360 to nearly flux maxima
after day 7360.
This phase jump causes the peak in the Fourier transform
to split into two, straddling the true peak.
By analogy with amplitude modulation
of a carrier signal, one should still see the carrier
in the power spectrum, with symmetric sidebands on either
side.  However, if the carrier (or any signal) 
experiences a phase jump, then its power will be split
into two frequencies, neither of which is the true one.
The frequency splitting of the signal should
be equal to the frequency with which the phase jumps,
in this case $1/14.7$~hr$^{-1}$, which is consistent
with the observed splitting of $1/14.0$~hr$^{-1}$.

We tested this interpretation by measuring the power spectrum
of the first part of the December 2 observation,
before the minimum of the beat cycle at day 7360 in
Figure~\ref{fig:cv_light_curve_1}.  This restricted power spectrum has
a single peak at 94.6 minutes (Figure~\ref{fig:xmm_power}b),
consistent with the period from October 20.  The same period is also
confirmed in a \chandra\ observation described below.
Therefore, we adopt the average value of \spin\
from the two \xmm\ observations as the probable
spin period of the WD.

We further confirm this interpretation using simulated light curves
and power spectra.  As an approximation of the light curve,
a squared sinusoid with a 94.8 minute period was amplitude modulated at
the 14.7~hr period, and a phase jump of varying angle was introduced
at day 7360.  This reproduced well the splitting of the spin signal in the
observed power spectrum.  The best match to the observed light curve
and power spectrum was achieved for a phase jump of $\approx120^{\circ}$,
with an uncertainty of $\sim20^{\circ}$.

We hypothesize that the WD spin modulates
the apparent X-ray flux by self occultation of the base of a
column that accretes onto a magnetic pole in an AM Her-like
system.  The 14.7~hr amplitude modulation superposed on this oscillation
occurs as the secondary star migrates in the rotating frame of the WD.
The $\approx120^{\circ}$ phase jump of the $94.8$ minute signal
at the minimum of the beat cycle corresponds to switching
of the accretion from one magnetic pole to the other.
Evidently the accretion rate drops almost to zero during the switch.
In this interpretation, 14.7~hours is one-third of the beat period
between the spin and the orbit; therefore, the orbital
period is \orbit, if it is longer than the spin period.
(Here we have estimated by eye from the light curve an uncertainty
of $\pm1.2~$hr on the 14.7~hr period.)
The 98.41 minute optical spectroscopic period is consistent with
the orbital period deduced from the X-ray light curve.

The OM light curves in Figures~\ref{fig:cv_light_curve_2}
and \ref{fig:cv_light_curve_1}, and MDM optical light curves in
Figure~\ref{fig:optphot}, display these effects as well, with
an amplitude that is smaller than the X-ray amplitude on the
spin period, but similar to the X-ray amplitude on the beat
period.  This supports the hypothesis that the emission is
from an accretion stream rather than a disk.  The optical
light may be less modulated by the spin of the white dwarf
than the X-rays, because it is coming from higher in the
accretion column.  But the varying accretion rate affects
the X-ray and optical luminosity equally on the beat period.

Notably, the X-ray hardness ratio shown in 
Figures~\ref{fig:cv_light_curve_2} and \ref{fig:cv_light_curve_1}
becomes softer in the dips.  This may indicate that a broader,
warm area of emission is present that is occulted less completely
than the central, hot base of the accretion column (see review
by \citealt{wic89}).  Alternatively, there may be residual
luminosity from another accreting region that is cooler
and less luminous.

\begin{figure*}
\centerline{
\includegraphics[angle=0,width=1.05\linewidth,clip=]{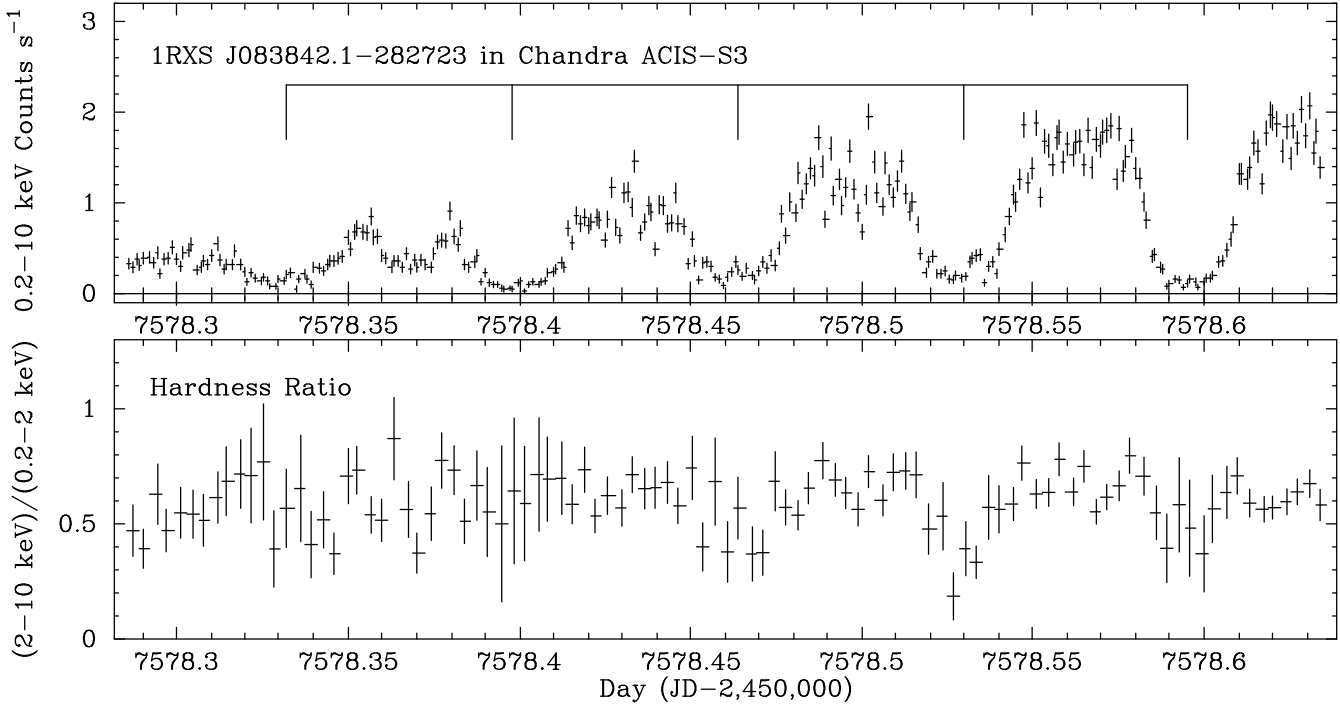}
}
\vspace{-1.8truein}
\caption{Top: Background-subtracted \chandra\ ACIS-S3 light curve of
\rxs, in 100~s bins.  For comparison the tick marks are at intervals
corresponding to the best-fitted period of 94.8 minutes in the \xmm\ data.
Bottom: Hardness ratio of counts in the
(2--10 keV)/(0.2--2 keV) bands, in 300~s bins.
}
\label{fig:chandra_light_curve}
\end{figure*}

\subsection{\chandra\ Analysis}

We fitted the \chandra\ X-ray spectrum with a thermal bremsstrahlung
model in XSPEC, finding a temperature of 11.2~keV (Figure~\ref{fig:rxs_spec}
and Table~\ref{tab:spectra}), consistent with the \xmm\ results.
X-ray photon arrival times were transformed to Barycentric Dynamical Time
and extracted from a $2^{\prime\prime}$ radius around \rxs.
Figure~\ref{fig:chandra_light_curve} shows the 0.2--10~keV
background subtracted light curve in 100~s bins.
Although observation is shorter than the \xmm\ ones, the behavior 
of \rxs\ is consistent with that observed previously.  The spin period and
the spin/orbit beating effect are both evident.  A peak at 93.9
minutes appears in the power spectrum (Figure~\ref{fig:xmm_power}d),
consistent with the \xmm\ values.
Since this short observation does not span a minimum of the beat
cycle, it does not clearly show a phase jump in the spin cycle.
However there is an ``extra'' dip at day 7578.37 in
Figure~\ref{fig:chandra_light_curve} that may be
due to partial pole switching of the accretion.

\begin{figure*}
\centerline{
\includegraphics[angle=0,width=1.05\linewidth,clip=]{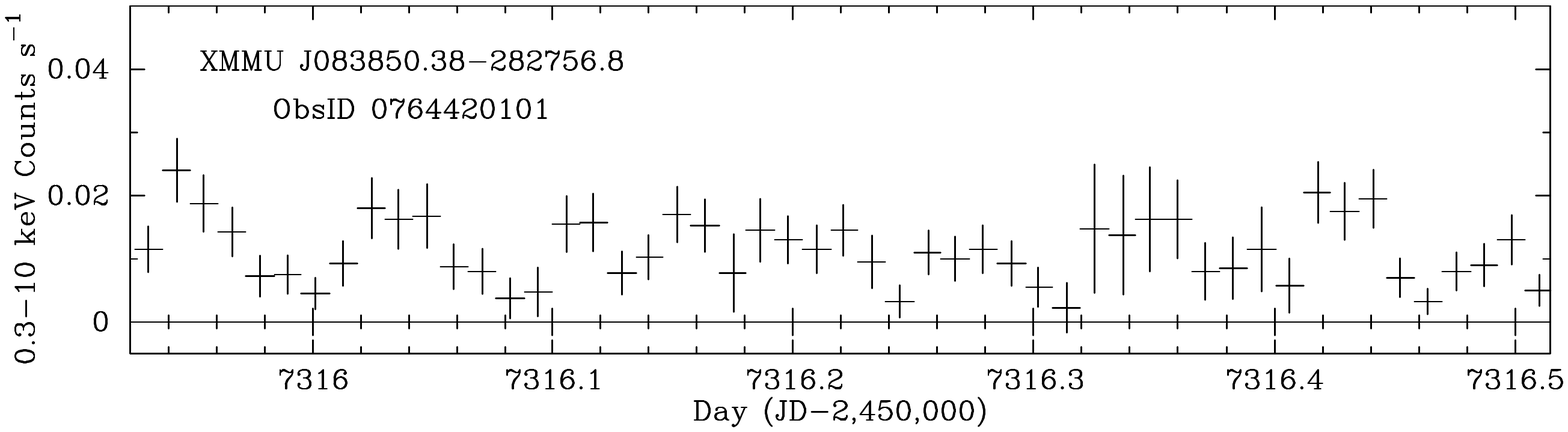}
}
\vspace{-3.75 truein}
\centerline{
\includegraphics[angle=0,width=1.05\linewidth,clip=]{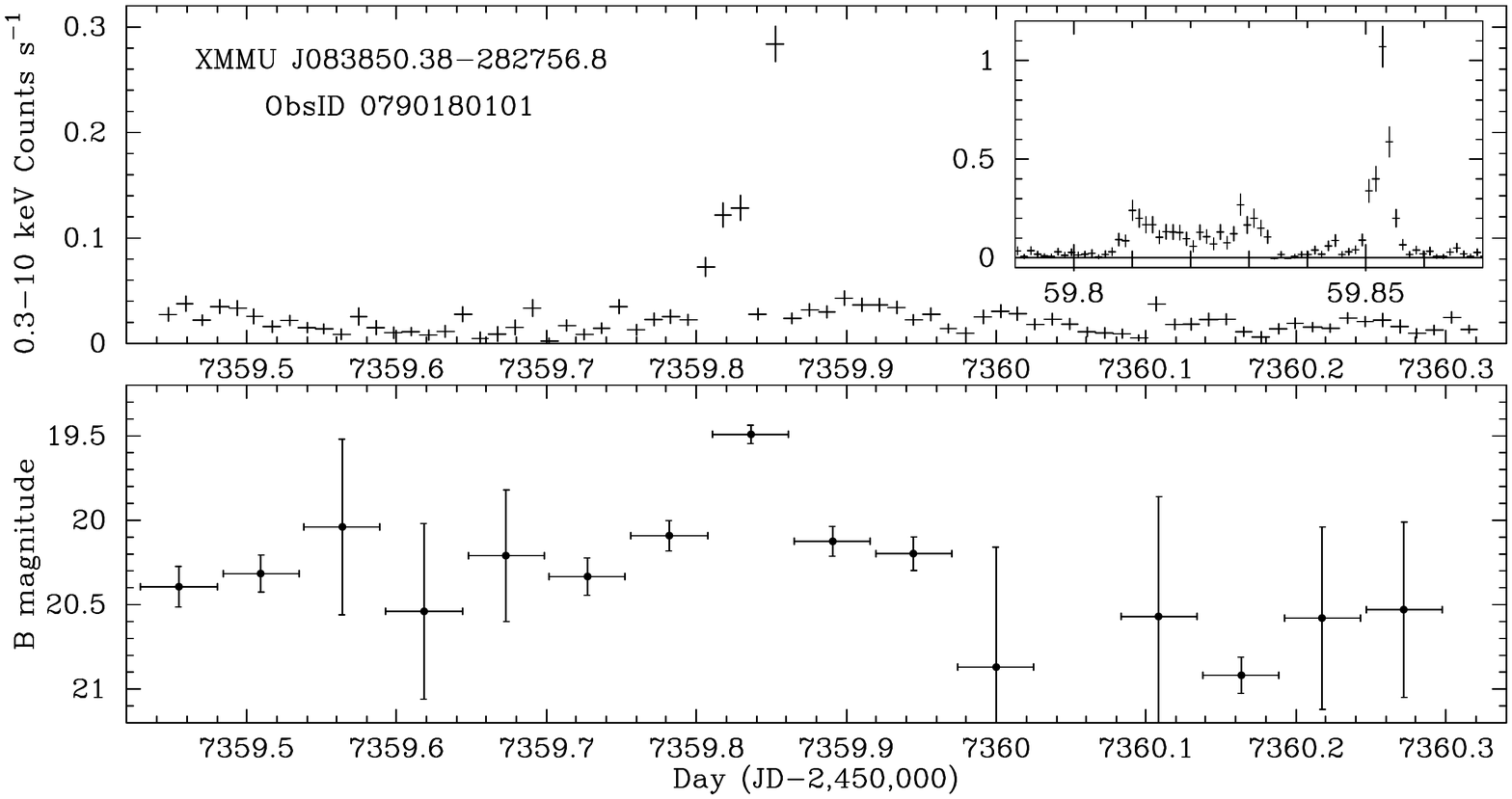}
}
\vspace{-1.7 truein}
\caption{Background-subtracted \xmm\ light curves of the MSP 
candidate \msp.
Top: EPIC pn 0.3--10~keV light curve in 1000~s bins
on 2015 October 20 (ObsID 0764420101).
Middle: EPIC pn 0.3--10~keV light curve in 1000~s bins
on 2015 December 2 (ObsID 0790180101).
The inset shows the 1.2~hr flaring episode at day 7359.8
at higher resolution (100~s bins).
Bottom: \xmm\ OM magnitudes from 4400~s exposures in $B$-band
on 2015 December 2 (ObsID 0790180101).
There is a gap in the data at day 7360.05.
}
\label{fig:xmm_light_curve_1}
\end{figure*}

\begin{figure*}
\vspace{-1.0in}
\centerline{
\includegraphics[angle=0,width=0.75\linewidth,clip=]{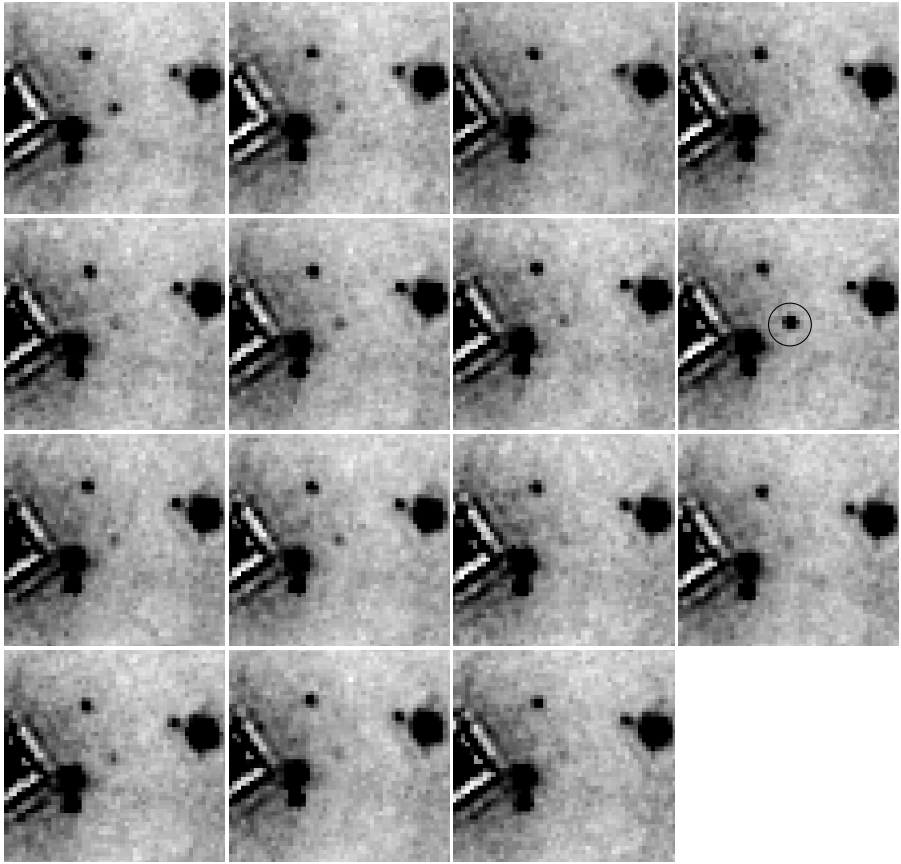}
}
\vspace{-1.1 truein}
\caption{\xmm\ OM $B$-filter image cutouts around the MSP candidate \msp\
from the 2015 December 2 observation (ObsID 079018010).
Time increases from left to right and top to bottom,
corresponding to the 15 points in Figure~\ref{fig:xmm_light_curve_1}.
The source is circled in the image coinciding with the X-ray flare.
The field is $50^{\prime\prime}\times50^{\prime\prime}$.
}
\label{fig:om_images}
\end{figure*}

\begin{figure}
\centerline{
\includegraphics[angle=0,width=1.15\linewidth,clip=]{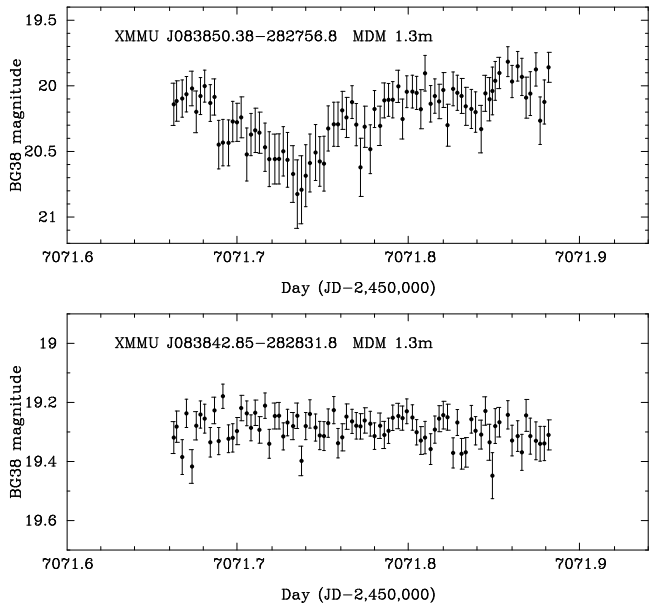}
}
\vspace{-1.7in}
\caption{MDM 1.3m light curves of source ``b'' (MSP candidate \msp, top)
and source ``c'' (QSO \qso, bottom), from the 2015 February 18 time series
(Table~\ref{tab:optlog}).  The 20~s images were combined in groups
of 10, for an exposure time of 200~s per point.
}
\label{fig:optbin}
\end{figure}

\begin{figure}
\centerline{
\includegraphics[angle=0,width=1.07\linewidth,clip=]{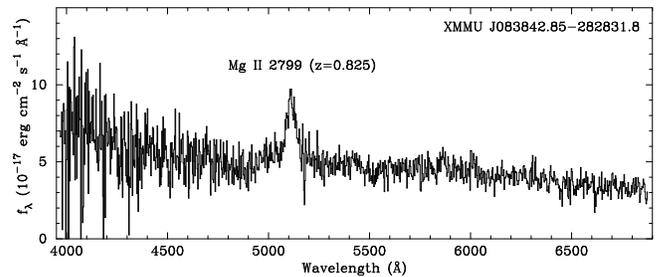}
}
\vspace{-1.15in}
\caption{Optical spectrum of \qso\ from OSMOS on the MDM 2.4m telescope.
The spectrum has been binned to 2.9\,\AA.
}
\label{fig:qsospec}
\end{figure}

\section{\msp}

Given the lack of association of any cataclysmic variable
with a persistent \fermi\ $\gamma$-ray source, and the absence
of a plausible physical mechanism connecting \rxs\ with \fgl,
we searched for another counterpart.
In this section we describe the analysis of source ``b,'' \msp.

\subsection{\xmm\ Analysis}

We extracted pn light curves of \msp. The MOS detectors,
operated in small window mode, did not cover this source.
Figure~\ref{fig:xmm_light_curve_1} shows a moderately variable
light curve, and an intense flaring episode on 2015 December 2 that
lasted $\approx1.2$~hr, with doubling times of $\sim100$~s
or less.  A power-spectrum analysis, excluding the
flare, does not indicate any periodicity.  Upper limits
on orbital modulation in the 0.5--10~hr period range are $\sim35$\%
for a sinusoidal amplitude.
The energy spectrum of \msp\ is well fitted by a hard power law of
photon index $\Gamma=1.5\pm0.1$ (Table~\ref{tab:spectra}).
A thermal bremsstrahlung model is also acceptable, but its
temperature is very high and poorly constrained ($14.4_{-5.2}^{+11.1}$~keV).
A blackbody fit is unacceptable, with reduced $\chi^2_{\nu}=2.6$ and
$N_{\rm H} = 0$.

On December 2, the OM
observed \msp\ for the entire span of the observation.
It revealed an optical counterpart for this source that is slowly
varying by $\sim1$~magnitude, and has a bright maximum
coincident with the X-ray flare, as shown
in Figures~\ref{fig:xmm_light_curve_1} and \ref{fig:om_images}.
(The OM was windowed on 2015 October 20 so that it did
not obtain continuous coverage of \msp.)

\subsection{Time-series Optical Photometry}

MDM 1.3m images from 2015 February 17 and 18 are the only
ones that cover this source.
Because of its faintness and the short individual exposure times,
we binned the images in groups of 10.  Due to weather, only the
2015 February 18 observation yields a useful light curve
(Figure~\ref{fig:optbin}).   It shows a broad minimum
that is characteristic of the heating light curve of a BW
companion to an MSP.  The X-ray and optical behavior
of \msp\ is similar to that of BW pulsar binaries identified with
\fermi\ sources, such as PSR J1311$-$3430, which has both
optical and X-ray flares \citep{rom12}.  While there is only one broad
minimum during the 5.4~hr span of the data, this is not
necessarily a lower limit on the orbital period because of
the possibility of flaring, which may mask a second dip
near the end of the time series.

\section{\qso}

A third X-ray source on the pn CCD, \qso, labeled ``c'' in the
images of Figure~\ref{fig:finder}, falls within the error
circle of \fgl.  Its $\Gamma=2.2$ power-law X-ray spectrum
and relatively steady optical flux in the OM 
and MDM 1.3m time series (Figure~\ref{fig:optbin})
suggest that it is a QSO.  The fitted $N_{\rm H} = 1.26\times10^{21}$~cm$^{-2}$ is
consistent with the total 21~cm $N_{\rm H}=1.4\times10^{21}$~cm$^{-2}$
on the line of sight \citep{kal05}.  This object was also
listed as a low-probability blazar candidate for \fgl\
by \citet{massaro13}, from its colors in the
{\it Wide-field Infrared Survey Explorer} data.
Lacking a radio detection in the NVSS, it is unlikely
to be a blazar.
We obtained an optical spectrum of \qso\ using the Ohio State
Multi-Object Spectrograph (OSMOS) on the MDM 2.4m telescope.
In the wavelength range 3960--6880 \AA\ it shows a
single broad emission line, which we identify as \ion{Mg}{2}
$\lambda2799$ at $z=0.825\pm0.001$, with an equivalent width
of $\approx45$~\AA\ (Figure~\ref{fig:qsospec}).
This confirms that it is a non-blazar AGN,
and probably unrelated to the \fermi\ source.

\section{Discussion}

\subsection{\rxs}

The X-ray light curve of \rxs, with its deep, broad dips at
a period of \spin, is typical of those polars in which
self occultation of the emitting magnetic pole by the WD is responsible
for the modulation.  It cannot be an eclipse by the orbiting secondary
star or the accretion stream, because either of those would be
much narrower.  Eclipse ingress or egress of the entire WD
would take only a few seconds, less than one bin of the light curves,
based on the orbital velocities estimated below.  The eclipse
duration would be $<0.1$ cycles given the radius of the secondary
star.  See, e.g., \citet{wor15} for a light curve of
an eclipsing polar.

The slower but equally deep
14.7~hr modulation requires yet another explanation.  
It also is too broad to be an eclipse by the orbiting secondary
or accretion stream.  The hardness ratio does not indicate
photoelectric absorption as a cause.
Instead, we hypothesize that the long period results
from interruption of the accretion stream due to an orbit
that is asynchronous with respect to the WD rotation.
This modulates the accretion rate onto the WD with a 14.7~hr period.

In addition, the spin cycle is thrown out of phase by $\approx120^{\circ}$
at the minimum of the 14.7~hr modulation.  This is interpreted
as switching of the accretion between not quite
antipodal regions,
when the accretion almost stops.  The effect on the power spectrum
is exactly analogous to amplitude modulation
of a carrier signal, with a phase jump, which splits
the signal into two periods straddling the true one.
We have recovered the true period,
which we interpret as the WD spin,
in the three X-ray observations reported here.
There is no direct manifestation of the orbit in the
power spectrum, but the 14.7~hr period most
likely represents the time it takes for the companion star
to migrate $\approx120^{\circ}$ around the WD, i.e.,
one-third of the beat period of the spin and the orbit.
The implied orbital period is \orbit,
assuming that it is longer than the spin period.

The emission-line radial velocity curve has a period of
\optper, which is consistent with the
X-ray inferred orbital period, and not with the spin.  But
the radial velocity amplitude of 274~km~s$^{-1}$ is too large to
be an orbital velocity of the WD, and the lines are
too broad, $\sim1500$~km~s$^{-1}$ to be coming from the
heated face of the secondary star.
A typical CV with orbital period of 100 minutes has
a $0.8\,M_{\odot}$ WD and a $0.1\,M_{\odot}$
secondary \citep{sav11}.  Typical orbital velocities
are then $\approx56$~km~s$^{-1}$ for the WD and
$\approx445$~km~s$^{-1}$ for the secondary.
On the other hand, free-fall velocity onto the WD
is up to 3900~km~s$^{-1}$.
So we conclude that the emission lines are located
high in the accretion column.

We also examined the EW of the H$\alpha$ emission line
as a function of spectroscopic phase.  The
EW varies $\approx0.5$ cycles out of phase with 
respect to the radial velocity (Figure~\ref{fig:specfold}).
The EW is therefore greatest at the point of maximum blueshift,
phase 0.75.  Assuming radial accretion, this is when the accreting pole is
on the far side of the WD and maximally occulted.  The emission-line
source could be higher up in the accretion column than the
optical continuum source, and therefore less modulated.
This would result in a larger EW at phase 0.75, where
the continuum is weakest.

The original group of four asynchronous polars
have spin and orbit periods which
periods differ by $\simle2\%$ (\citealt{sch07}, and references
therein).  In \rxs, the difference we infer is $3.7\%$.
A fifth asynchronous polar has
been proposed to explain the post-nova light curve
of V4633~Sgr \citep{lip08}, which had transient period that
decreased from 185.6 minutes to 183.9 minutes
in addition to a stable period of 180.8 minutes.
The authors attributed the longer period to the spin period,
which was perturbed by the outburst, while the shorter
period is orbital.  It is generally observed that asynchronous polars
are evolving toward synchronization on a short time scale, which
suggests that nova eruptions may be the cause of all asynchronous
polars.

Other CVs with more complex light curves and
multiple periods have also been proposed as extreme asynchronous
polars, most notably RX J0524+42, dubbed ``Paloma'' \citep{sch07,jos16},
and IGR~J19552+0044 \citep{ber13,tho13}.  Paloma is suggested
to have an orbital period of 157 minutes, and a spin period
that is either 136 or 146 minutes, which are asynchronous
by 14\% or 7\% respectively \citep{sch07}.  IGR~J19552+0044
has a spectroscopic period of 1.39~hr that disagrees with
its photometric period of 1.36~hr \citep{tho13}, while
\citet{ber13} finds X-ray periods of 1.38~hr and 1.69~hr.
For these systems with highly discrepant periods, it is
not yet known if they are evolving toward synchronization, but
if so, it may be for the first time.
Among this group, \rxs\ has the simplest light curve and power spectrum,
and thus may be a key to interpreting the other, more complex cases.

\subsection{\msp}

None of the properties of the CV \rxs\ suggest a connection
with \fgl.  Using machine learning algorithms, \citet{mir16}
classified \fgl\ as a high-confidence pulsar based on its
$\gamma$-ray spectral shape and variability index.  Accordingly,
we propose \msp\ as the millisecond pulsar counterpart of \fgl.

X-ray properties of BW pulsars were reviewed by \cite{gen14}
and \citet{aru15}.  They have hard power-law spectra and
sometimes a soft thermal component from the neutron star
surface. The non-thermal emission is thought to be synchrotron
from an intrabinary shock between the pulsar wind and the
companion stellar wind.  In some cases the X-ray flux is
modulated on the orbital period, due either to relativistic
beaming, or self occultation of the emitting region near the
surface of the companion star \citep{rom16}.

The optical light curves of BWs are modulated by photospheric
heating.  Usually the heated face of the substellar 
companion is much brighter than the cool photosphere on the
``night'' side.  In some cases, the heating light curve is
not symmetric with respect to the line between the stars,
which could be due to an asymmetric shock, or to channeling
of the pulsar wind by intrinsic magnetic fields on the companion
\citep{tan14,li14}.
In PSR J1311$-$3430, bright flares have been seen in optical
and X-rays \citep{rom12}, which could be coming from the
companion's magnetic fields.  PSR J1311$-$3430 is a short-period (94 minute)
system, which, being in tidally locked rapid rotation, could
enhance the coronal magnetic field \citep{rom15}.

The hard X-ray power-law spectrum and X-ray and optical flare
seen in the \xmm\ observation of \msp\ strongly
motivate its identification as the MSP counterpart of \fgl.
Typical \fermi\ MSPs are at distances of $\sim1$~kpc.
If so, its X-ray luminosity would be $2\times10^{31}$ erg~s$^{-1}$,
in the range of both BWs and redbacks in the radio pulsar state
\citep{rob15}.  In contrast, accreting redbacks have
$L_x \sim 3\times10^{33}$ erg~s$^{-1}$ \citep{bog15a}.

This would be the first time that a simultaneous X-ray
and optical flare is seen from a BW. It may be a
short-period system similar to PSR J1311$-$3430, but the
X-ray light curve doesn't reveal an orbital period.
Any orbital modulation in the 0.5--10~hr period range
has $<35$\% amplitude for an assumed sinusoid.
The cadence of the 4400~s OM exposures is too
long to test for such a period, which may, in addition,
be masked by the flaring behavior.  The MDM optical data,
on the other had, have adequate cadence, and a dip that
is characteristic of a heating light curve, but the
dip does not repeat within the 5.4~hr time series
(Figure~\ref{fig:optbin}), which suggests that the
period is $>3.4$~hr.  Alternatively, a second dip
may be masked by a flaring episode.

BW optical light curves are usually modulated by several magnitudes,
while the dip in \msp\ is only $\approx0.8$ magnitudes.  This could
mean that the inclination angle of the binary is small, or that the
companion is a low-mass main sequence star,
i.e., a redback, which is brighter than a BW.  Optical flares
have also been seen from the redback PSR J1048+2339 \citep{den16}.
Additional time-series
photometry, optical spectroscopy, and a radio or $\gamma$-ray
pulsar detection, could confirm the identification,
and resolve the remaining questions about
the basic parameters of the binary system.

\section{Conclusions}

The $\gamma$-ray properties of \fgl\ have suggested that it
is a pulsar.  In this X-ray and optical study,
we first concluded that \rxs, the brightest
X-ray source in its error circle, is an unusual CV that falls
there by chance.  It has a simple X-ray light
curve that is strongly modulated at two periods,
with a phase jump that indicates pole-switching
of the accretion in an asynchronous polar.
The spin is manifest as a 94.8 minute X-ray period
caused by self occultation of the accreting pole,
while the switching interval is 14.7~hours.  This implies
an orbital period of 98.3 minutes, consistent with
the 98.41 minute optical spectroscopic period.
The strong optical
and X-ray modulation on the 14.7~hr period can be explained
by nearly complete interruption of the accretion stream
as it switches poles.  A dedicated optical monitoring
campaign could obtain a more precise value for the
beat period, test the proposed geometry of the accretion spots
in more detail, and determine whether the spin
and orbit are evolving toward synchronism.

X-ray and optical observations identify
a second, highly variable object in the error circle of \fgl.
\msp\ is modulated on a time scale of hours in the optical,
in addition to having shown one simultaneous X-ray and optical flare.
It has a hard, nonthermal X-ray spectrum.  These properties are
compatible with black widow or redback millisecond pulsar systems
that have been discovered as counterparts of \fermi\ sources.
A binary period, an important test of this hypothesis,
is not yet revealed by the available data,
but follow-up time-series photometry, optical spectroscopy,
and/or a radio pulsar detection, should be able to determine
the orbital parameters of the system.

\section{Acknowledgements}

We thank Eric Alper for obtaining time-series photometry
of \rxs\ in 2016 March, and Jessica Klusmeyer
for the optical spectrum of the QSO \qso.
The MDM Observatory is operated by Dartmouth College,
Columbia University, the Ohio State University, Ohio University,
and the University of Michigan.
The results reported in this article are based in part on
observations made by the \chandra\ X-ray Observatory.
Support for this work was provided by the National Aeronautics and
Space Administration through \chandra\ Award Number SAO GO6-17027X
issued by the \chandra\ X-ray Observatory Center,
which is operated by the Smithsonian Astrophysical Observatory
for and on behalf of the National Aeronautics Space Administration
under contract NAS8-03060.
This investigation also uses observations obtained with \xmm,
an ESA science mission with instruments and contributions directly
funded by ESA Member States and NASA.

\end{document}